\documentclass[a4paper,11pt,dvipsnames,final]{article}
\usepackage{import}
\usepackage{Preamble}

\preprint{BRX-TH-6732}

\title{\boldmath Quantum covariant bit threads}

\author[a]{Matthew Headrick,}
\author[b]{Sreeman Reddy Kasireddy,}
\author[c]{and Andrew Rolph}

\affiliation[a]{Martin Fisher School of Physics, Brandeis University, Waltham MA 02453, USA}
\affiliation[b]{Department of Particle Physics and Astrophysics, Weizmann Institute of Science, Rehovot 7610001, Israel}
\affiliation[c]{Vrije Universiteit Brussel (VUB) and The International Solvay Institutes, Pleinlaan 2, B-1050 Brussels, Belgium}
\emailAdd{headrick@brandeis.edu}
\emailAdd{sreeman@weizmann.ac.il}
\emailAdd{andrew.d.rolph@gmail.com}

\abstract{
Entanglement entropy plays a central role in holography. The quantum extremal surface (QES) formula is the standard tool in settings where both dynamics and bulk quantum corrections are important, such as in black hole evaporation. We derive and explore QES-equivalent prescriptions --- quantum covariant bit threads --- that formulate holographic entanglement entropy in terms of minimal or maximal bulk flows rather than extremal surfaces. Like classical covariant bit threads, our prescriptions come in two types: (typically) spacelike V-flows, and timelike U-flows. To account for bulk quantum corrections, the flow lines can start and end at points in the bulk, with the distribution of endpoints controlled by bulk entanglement entropies. In our bit thread language, the QES is a quantum bottleneck to the flow, with some threads passing through it (accounting for the area term in the generalized entropy) and others jumping across it (accounting for the entropy term). Our prescriptions provide a complementary perspective to the QES formula and offer new insights on important phenomena such as black hole evaporation, islands, and baby universes. \\
\\
A video abstract is available \href{https://youtu.be/8U8VdLAIku0}{here}.
}

\makeatletter
\gdef\@fpheader{}
\makeatother
\begin{document}
\nolinenumbers
\maketitle
\flushbottom

\section{Introduction}
\label{sec:intro}

The Ryu-Takayanagi (RT) formula \cite{Ryu:2006bv} and its generalizations have arguably been some of the most fruitful and far-reaching discoveries in quantum gravity in this century. 
The RT formula gives the entanglement entropy $S(A)$ of a boundary spatial region $A$ in terms of the area of a certain bulk minimal surface. The formula has a ``bit thread" reformulation in terms of the maximal flux through $A$ of a divergenceless, norm-bounded bulk vector field, or equivalently the maximal number of Planck-area bulk threads connecting $A$ to its complement \cite{Freedman:2016zud}:
\bne \begin{split}  S(A) &= \min_{\gamma \sim A} \left(\frac{\text{Area}(\gamma)}{4G_N}\right)\\
&= \max_v \int_A *v \quad\text{ subject to } d*v = 0,\, |v| \leq \frac{1}{4G_N}. \label{eq:introRTBIT} \end{split} \ene
The second equality in  eq.~\eqref{eq:introRTBIT} is a continuum version of the max-flow/min-cut theorem. A loose analogy for this equivalence is that the area of the smallest cross section of a pipe and its maximum flow rate are the same.

A bit thread prescription is any formula for calculating the entropy $S(A)$ using flows or curves in the bulk, rather than surfaces. Eq.~\eqref{eq:introRTBIT} is an example, where the eponymous bit threads are the flow lines of $v$. Bit threads offer certain conceptual and technical advantages over the original minimal-surface formulation~\cite{Freedman:2016zud}, such as making the physical meaning of entanglement entropy inequalities manifest, and being convex optimisation problems.

The RT formula, and its bit thread equivalent, are only applicable under certain conditions: (1) the state should be time-reversal invariant, and (2) the bulk should be governed by classical Einstein gravity. There are important generalisations of the RT formula for when these two conditions are relaxed: first, the covariant Hubeny-Rangamani-Takayanagi (HRT) formula \cite{Hubeny:2007xt}, which allows for time-dependence, and second, the quantum extremal surface (QES) formula \cite{Engelhardt:2014gca},
which allows for both time-dependence and bulk quantum corrections. 
The significance of these two generalizations was definitively established by their use in the derivation of the Page curve for the entropy of an evaporating black hole \cite{Penington:2019npb,Almheiri:2019psf}, a landmark result in the quest to resolve the puzzles of quantum gravity.

\begin{figure}[htbp]
    \centering
    \includegraphics[width=0.8\linewidth]{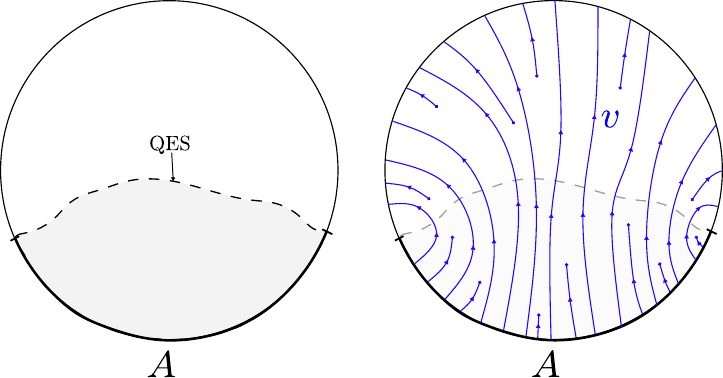}
    \caption{A time-reflection symmetric AdS time slice and two different but equivalent ways of calculating the boundary entropy $S(A)$: either minimising the generalised entropy of surfaces homologous to $A$ (the QES prescription), or maximising the flux of a constrained vector field $v$ through $A$ (the quantum bit thread prescription).}
    \label{fig:QBitThreadsIntro}
\end{figure}

\begin{figure}[htbp]
    \centering
    \includegraphics[width=0.9\linewidth]{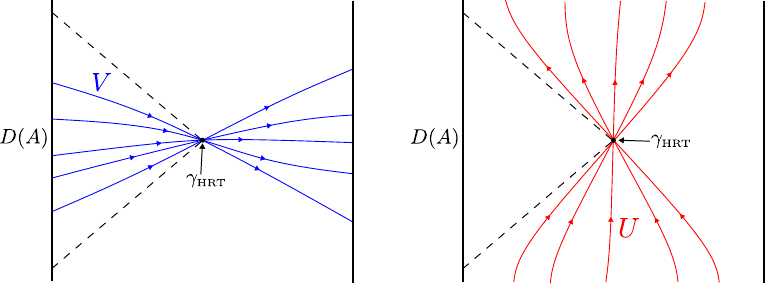}
    \caption{Classical V and U-flows in an asymptotically AdS spacetime. In V-flow prescriptions, one maximises the flux of a (typically) spacelike constrained vector field $V$ through the boundary domain of dependence $D(A)$. In U-flow prescriptions, one minimises the flux of a timelike constrained vector field $U$ through past infinity. The prescriptions are equivalent to each other and the HRT formula.
    }
    \label{fig:U_VThreads2}
\end{figure}

An important aspect of the bit thread programme has been to find corresponding generalisations for when the above two conditions are relaxed. References for the current state-of-the-art with regards surface-based versus bit thread-based prescriptions are collected in Table~\ref{tab:HEEst}. 
So far, bit thread formulas equivalent to HRT and the QES formula applied to static states have been found~\cite{Headrick:2022nbe,Agon:2021tia,Rolph:2021hgz,Headrick:2025awv}. 
We will review the technical details of these prescriptions in Sec.~\ref{sec:CQCBT}, but, in short: 
\begin{enumerate} 
\item In the static quantum bit thread formulas, bulk quantum corrections to $S(A)$ are captured by modifying the divergence constraint and so allowing threads to start and end at bulk points, with the net number bounded by each region's entropy. This allows threads to effectively jump over the geometric bottleneck; see Fig.~\ref{fig:QBitThreadsIntro}. 
\item In the classical covariant formulas, threads are allowed to move in the full spacetime, rather than just through a static slice. There are also now \textit{two} complementary thread types: U and V-threads, with the extra type essentially coming from the inclusion of the time direction. V-threads are (typically) spacelike, and we maximise their flux through $D(A)$. U-threads are timelike, and we minimise their flux through the past boundary. See Fig.~\ref{fig:U_VThreads2}.
\end{enumerate}

\begin{table}[htbp]
\centering
\resizebox{\textwidth}{!}{%
\begin{tabular}{cc|cc}
    & & Classical & Quantum\\
    \hline
    \multirow{2}{*}{\textbf{Surface-based}} & Static & RT~\cite{Ryu:2006bv} & (Static) QES~\cite{Engelhardt:2014gca} \\
    & Covariant & HRT~\cite{Hubeny:2007xt} & (Covariant) QES~\cite{Engelhardt:2014gca} \\
    \hline
    \multirow{2}{*}{\textbf{Bit threads}} & Static & Classical static bit threads~\cite{Freedman:2016zud} &
    Quantum static bit threads~\cite{Rolph:2021hgz,Agon:2021tia,Headrick:2025awv} \\
    & Covariant & Classical covariant bit threads~\cite{Headrick:2022nbe} &
    \textit{Quantum covariant bit threads} \\
\end{tabular}%
}
\caption{The surface-based prescriptions of holographic entanglement entropy and their bit thread equivalents. Filling in the gap in the bit thread prescriptions --- quantum covariant bit threads --- is the purpose of this paper.}
\label{tab:HEEst}
\end{table}

The core of this paper is the derivation and exploration of new bit thread formulas that are equivalent to the fully covariant QES formula: \textit{quantum covariant bit threads}. Doing so completes Table~\ref{tab:HEEst} and an important piece of the bit thread programme: a bit thread equivalent of every surface-based prescription for holographic entanglement entropy in Einstein gravity.%
\footnote{Bit thread formulas for generalizations of the RT formula in other directions have also been explored, including stringy corrections \cite{Harper:2018sdd}, entanglement of purification \cite{Harper:2019lff}, non-AdS holography \cite{Du:2024xoz,Gupta:2025jlq}, and multi-party entanglement \cite{Harper:2021uuq,Harper:2022sky,Headrick:2020gyq}.
}

Our new quantum covariant prescriptions combine features of both the static quantum and classical covariant formulas.
Similarly to the classical covariant prescriptions, there are both U- and V-threads, and these obey the same constraints as their classical cousins, except for the divergence constraint. Similarly to the quantum static threads, the divergence constraint is such that the net number of threads that can enter or leave a region is bounded by the region's entropy, except now covariantised to spacetime regions. There is no static analogue of U-threads, which makes finding their quantum formulation more challenging, because the static quantum prescriptions of~\cite{Rolph:2021hgz, Headrick:2025awv} are less of a guide.

As with all bit thread formulas, the threads have considerable freedom, and there are
many optimal thread configurations, but all of them share the feature that the threads are collimated and maximally (for the V-threads) or minimally (for the U-threads) packed at the QES. There is both a geometric ``classical" bottleneck for quantum threads, which accounts for the area piece of the generalised entropy, and, through the divergence constraints, a non-local, ``quantum" bottleneck to threads effectively jumping across the geometric bottleneck.

Our new bit thread prescriptions allow for discussion and investigation that is complementary to those based on
the QES formula, using thread configurations rather than extremal surfaces.
We will use our prescriptions to study the behaviour of bit threads in settings where both the dynamics and quantum corrections are important, such as an evaporating black hole. 
We will find interesting phenomena, such as how entanglement islands manifest in bit thread language. 

There are various directions in which our results could be built upon, such as understanding entanglement entropy in non-AdS holography.
It is plausible that U-flow prescriptions in AdS/CFT could point towards natural proposals in dS holography, because U-threads are anchored on spacelike boundaries in direct analogy with how V-threads are anchored to the timelike AdS conformal boundary.

The paper is organised as follows. In Sec.~\ref{sec:CQCBT}, we review the older bit thread prescriptions, on which the new quantum covariant bit thread prescriptions of Sec.~\ref{sec:QCBIT} are built. In Sec.~\ref{sec:OptFlows}, we discuss features of optimal flows in the new prescriptions. We save our technical derivations for Sec.~\ref{sec:FlowsDerivs}, which starts with new surface-based prescriptions equivalent to the QES formula, including one that is concave-convex in its variables. This is a precursor to using these prescriptions to derive our quantum covariant bit thread formulas via Lagrange dualisation.

\section{Review and background}
\label{sec:CQCBT}

To set the stage for our new quantum covariant bit thread prescriptions, we will briefly recall the previous formulas upon which they are built: classical static, quantum static, and classical covariant. We start from the simplest and build upwards. Each formula will be expressed as a concave or convex optimization problem.

A bit thread is a constrained curve in the bulk spacetime. Bit thread prescriptions fall into two categories:
\begin{enumerate}
    \item \textit{Flow}-based, where we optimise over a constrained vector field, and the integral curves of the vector field are the bulk curves that we call bit threads. Flow lines and bit threads are synonymous in flow-based prescriptions.
    \item \textit{Thread distribution}-based, where we optimise over thread distributions, which are measures on sets of bulk curves. In these prescriptions, the bulk curves (i.e.\ threads) are in some sense the fundamental object; they are not the flow lines of some underlying vector field.
\end{enumerate}

In the following, we will only review flow-based prescriptions, for brevity. For their equivalent thread distribution prescriptions, and for more complete discussions, see the original papers~\cite{Freedman:2016zud,Headrick:2022nbe,Rolph:2021hgz,Headrick:2025awv}. 

\subsection{Static bit threads}

The RT formula and static bit thread prescriptions make the simplifying assumption of time-reflection symmetry: that there exists a boundary Cauchy slice containing $A$ on which the state is time-reflection symmetric. Then there exists a privileged, time-reflection symmetric bulk slice $\sigma_{\textrm{sym}}$. Both the surface-based and bit threads prescriptions simplify with the time-reflection symmetry assumption; one need only consider surfaces and flows contained within $\sigma_{\textrm{sym}}$.

\subsubsection{Classical}

Let $v$ be a 1-form (or covector field) on $\sigma_{\textrm{sym}}$, with $*v$ the Hodge dual $(D-2)$-form (with $D$ the spacetime dimension of the bulk; here $*$ is defined with respect to $\sigma_{\textrm{sym}}$).
Then,
\be
\text{\textbf{Classical static flow:}}\quad
S(A)=\sup\int_A*v \quad \text{over $v$ subject to:}
\quad|v|\le\frac1{4\GN}\,,
\quad d*v=0\,.
\label{eq:classs}
\ee
The first constraint is the norm bound (which imposes an upper bound on the density of threads), while the second is the divergenceless constraint, which expresses the fact that classical bit threads are continuous curves.

\subsubsection{Quantum}

For quantum static threads, we relax the divergenceless constraint, allowing the threads to start and end at points in the bulk. The quantum static flow prescription has two versions for its divergence constraint, which differ by a relative tightening. \\

\noindent\textbf{Quantum static flow:}%
\begin{align}
S(A)
= &\sup\int_A *v  \quad 
\text{over $v$ subject to: }
\nonumber\\[-2pt]
&
|v| \leq \dfrac{1}{4G_N} 
\quad \text{and 
}\quad \begin{cases}
\forall\,\tilde r\in\mathscr R_A^{\rm (sym)}\,,\quad
 -\displaystyle\int_{\tilde r} d*v
\leq S_{\rm b}(\tilde r)
& \quad\text{(loose)}
\\[12pt]
\forall\,\tilde r\in\mathscr R^{\rm (sym)}\,,\quad \left|\displaystyle\int_{\tilde r} d*v\right|
\leq S_{\rm b}(\tilde r)
&\quad\text{(strict)}\,.
\end{cases}
\label{qstaticflow}
\end{align}
Here $\sR^{(\textrm{sym})}$ is the set of all subregions of the bulk slice $\sigma_{\textrm{sym}}$, and $\mathscr{R}_A^{(\textrm{sym})} \subset \sR^{(\textrm{sym})}$ is the set of homology regions for $A$:
\be
\mathscr{R}_A^{(\textrm{sym})} =\left\{\tilde r\in\sR^{(\textrm{sym})}\,\big|\,\tilde{r} \cap \del \sigma_{\textrm{sym}} = A\right\}\,.
\ee
$S_{\rm b}$ is the bulk entropy, a non-negative function on $\sR^{(\textrm{sym})}$ that is assumed to obey the strong subadditivity (SSA) inequality, but is otherwise arbitrary. 
The strict divergence constraint is stricter in two ways: (1) the absolute value constrains both signs of the integrated divergence, and (2) the constraint is applied to the full set of bulk subregions rather than just the homology regions for $A$. 
Tightening the divergence constraint reduces the set of allowed flows, but does not affect the maximal flux through $A$; this non-trivial fact relies on the assumption that $S_{\rm b}$ obeys strong subadditivity.

\subsection{Classical covariant bit threads}
\subsubsection{Spacetime setup} \label{sec:STSTP}

\begin{figure}[h!]
    \centering
    \includegraphics[width=0.6
    \linewidth]{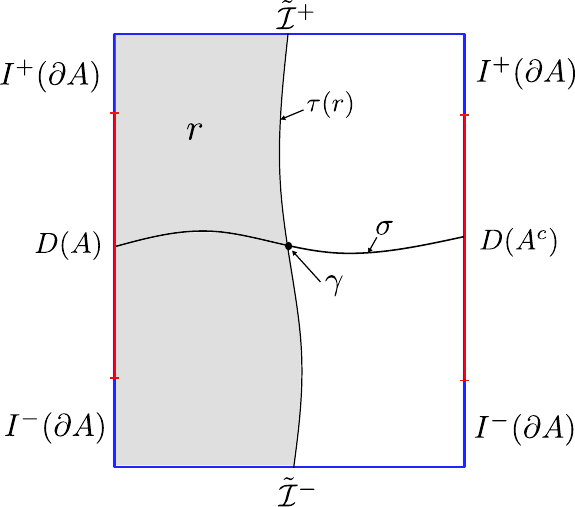}
    \caption{An illustration of the spacetime submanifolds defined in Sec.~\ref{sec:STSTP}. The figure shows a typical configuration, but the timesheet $\tau(r)$ can end anywhere on $\cI^\pm$ (including $I^\pm (\del A)$), and the slice $\sigma$ can end on $\tilde{I}^\pm$, because it is a level set of a global time function, not a Cauchy slice. We will not show $I^\pm (\del A)$ in later diagrams, to reduce clutter.
    }
    \label{fig:spacetime}
\end{figure}

Before we give the covariant bit thread prescriptions, 
we first have to describe our spacetime geometry and define some terminology. Fig.~\ref{fig:spacetime} is a visual guide to these details.

We denote the bulk spacetime by $M$, and assume that it is locally asymptotically AdS. In the classical covariant bit threads paper \cite{Headrick:2022nbe}, the bulk spacetime was assumed to be AdS-globally hyperbolic. Here, since we want to consider spacetimes like a completely evaporating black hole that contain Cauchy horizons, we will need to be more flexible. We therefore replace the condition of global hyperbolicity with that of \emph{stable causality}, which means that $M$ admits a global time function (a real function whose gradient is everywhere future-directed timelike). Any level set of a global time function is necessarily an acausal hypersurface (embedded codimension-1 submanifold). We will denote by \emph{slice} a hypersurface that is a level set of some global time function. If $M$ is globally hyperbolic, then a Cauchy slice is a slice; but a slice is not necessarily a Cauchy slice (e.g.\ a slice may end on a past or future singularity). We assume that any region $r$ of any slice $\sigma$ has an entropy $S_b(r)$, and that the entropies of the regions of $\sigma$ obey strong subadditivity.

The boundary of $M$ consists of a conformal boundary $N$, which we assume to be globally hyperbolic, possibly along with future and past boundaries $\tilde{\mathcal{I}}^\pm$: $\partial M=N\cup\tilde{\mathcal{I}}^+\cup\tilde{\mathcal{I}}^-$.%
\footnote{The boundary of $M$ can also include end-of-the-world branes (timelike boundaries at finite distance), and these can easily be accommodated in all of the flow formulas discussed in this paper, simply by constraining the flows to have vanishing flux through the brane. So, to avoid cluttering the formulas, we assume that $M$ has no end-of-the-world branes.}
Examples of future and past boundaries include black and white hole singularities. 
Given a boundary spatial region $A$, we denote by $D(A)$ its boundary domain of dependence, $D(A^c)$ the complementary domain of dependence (the domain of dependence of the complement of $A$ on any boundary Cauchy slice), and $\partial A$ the entangling surface on which they meet (which may be empty). With $D$ the spacetime dimension of $M$, $A$ has dimension $D-2$, $D(A)$ and $D(A^c)$ have dimension $D-1$, and $\partial A$ has dimension $D-3$.\footnote{The classical covariant bit thread paper \cite{Headrick:2022nbe} inserted a buffer between $A$ and $A^c$, in order to regulate UV divergences, and worked in the so-called regulated spacetime, the joint entanglement wedge of $A$ and $A^c$. Here we will not be so careful about using a UV regulator, and we will work directly in the full spacetime.} 

There is a decomposition of the spacetime boundary $\del M$ that will be useful.
Denote the latest and earliest Cauchy slices of $D(A)\cup D(A^c)$ as $\Sigma_\pm (A)$.
The conformal boundary $N$ can be decomposed into the domains of dependence $D(A)$ and $D(A^c)$, and the future/past regions $I^+ (\Sigma_+(A))$ and $I^- (\Sigma_- (A))$. The full boundary $\partial M$ can thus be decomposed into four regions, 
\bne \del M = D(A) \cup D(A^c) \cup \mathcal{I}^+ \cup \mathcal{I}^- \ene
where we define%
\footnote{In the spacetime setup as described so far,~\eqref{eq:calId} is equivalent to 
\bne \mathcal{I}^\pm =\tilde{\mathcal{I}}^\pm\cup I^\pm (\del A)\,. \ene
However, in Sec.~\ref{sec:V_BHs}, in the context of evaporating AdS black holes, we will extend $M$ by appending a ``bath" region, and~\eqref{eq:calId} is the correct general definition.
}
\bne \mathcal{I}^\pm:=\tilde{\mathcal{I}}^\pm\cup I^\pm (\Sigma_\pm (A)) \,. \label{eq:calId} \ene

We denote by $\mathscr{S}$ the set of all slices of $M$ whose intersection with $\del M$ is a boundary Cauchy slice for $N$, and $\mathscr{S}_A$ the set of slices whose intersection with $\del M$ is a boundary Cauchy slice for $D(A)\cup D(A^c)$. 
We denote by $\sR$ the set of codim-0 spacetime regions $r \subseteq M$ whose boundary component $\tau(r) := \del r \backslash \del M$ is piecewise timelike or null. We call $\tau(r)$ the \textit{timesheet} for $r$, and the set of all timesheets for $\sR$ is denoted by $\mathscr{T}$.%
\footnote{Note that the sets of timesheets $\tau$ and homology regions $r$ are not in one-to-one correspondence. 
This is because two different regions $r$ can have the same timesheet $\tau(r)$; the map $r\mapsto \tau(r)$ is surjective but not always injective.
For example, suppose that $M$ contains a disconnected closed baby universe $\cB \subset M$. Then, if $r$ is a homology region for $\tau$, then so is $r \cup \cB$. This distinction between the sets of timesheets and homology regions is not important for classical holographic entanglement entropy prescriptions (flow or surface-based) because only surface areas contribute at leading order in $N$ (and Area$(\sigma \cap \tau (r_1))$ = Area$(\sigma \cap \tau (r_2))$ if $\tau(r_1) = \tau(r_2)$), whereas the distinction does matter when including finite $N$ quantum corrections, because $\tau(r_1) = \tau(r_2)$ does not imply that $\Sgen(\sigma \cap r_1) = \Sgen(\sigma \cap r_2)$. This is why our prescriptions must use measures on $r$ rather than $\tau$, whereas the classical covariant bit thread prescriptions can use measures on either set.}
Timesheets can be thought of as the timelike counterpart to time slices.
Let $\sR_A \subset \sR$ be the set of spacetime homology regions for $D(A)$, relative to $\mathcal{I}^+\cup\mathcal{I}^-$:
\be\label{Rdef1}
\sR_A:=\{r \in \sR :D(A)\subseteq\partial r,D(A^c)\cap\partial r=\varnothing\}\,.
\ee
 For $r \in \sR_A$, $\tau(r)$ is homologous to $D(A)$ relative to $\mathcal{I}^+\cup\mathcal{I}^-$. The set of timesheets for $\sR_A$ is denoted $\mathscr{T}_A$.
Given $r\in\mathscr{R}_A$ and $\sigma\in\mathscr{S}_A$, the intersection $r\cap\sigma$ is a codimension-1 achronal region with entropy $S_{\rm b}(r\cap \sigma)$. 
Any codimension-2 surface $\gamma$ of the form $\tau(r)\cap\sigma$, where 
$\tau\in\mathscr{T}_A$ and $\sigma\in\mathscr{S}_A$, is homologous to $A$ and is therefore a candidate RT/HRT/QES surface. The QES formula equates $S(A)$ with the generalised entropy of the bulk region which is the minimal extremum of the generalised entropy functional,
\bne S(A) = \min \ext\, \Sgen(r \cap \sigma)\,. \label{eq:QESform} \ene
The extremisation is over codimension-1 spatial regions of the form $r\cap \sigma$, with $r\in\sR_A$ and $\sigma\in\sSA$. $\Sgen$
is the generalised entropy functional for Einstein gravity:%
\footnote{For gravitational EFTs with higher-curvature corrections (e.g. Gauss-Bonnet gravity), the QES formula is essentially unchanged; one simply adds (known) corrections to $\Sgen$'s area component~\cite{Camps:2013zua, Dong:2013qoa}. For static classical bit threads, there are corrections to the norm bound~\cite{Harper:2018sdd}. We expect something similar for quantum covariant bit threads, that higher curvature corrections can be accounted for in the norm bound, but we leave the derivation for future work.}
\bne \Sgen (r \cap \sigma) = \frac{\text{Area}(\tau(r) \cap \sigma)}{4G_N} + S_b (r \cap \sigma)\, .\ene

In Sec.~\ref{sec:OptFlows}, in the context of bit threads in evaporating AdS black hole backgrounds, we will generalise the spacetime setup discussed here to allow for open AdS boundary conditions through coupling to bath regions. Then not every AdS boundary Cauchy slice of the causal development of $A$ necessarily has the same entropy. In practice, the main change to our formulas is the replacement of $D(A)$ with $A$.

\subsubsection{Prescriptions}
Now we relax the time-reflection symmetry assumption and give the covariant bit thread prescriptions. 

There are two types of covariant bit thread prescription: one using V-flows and one using U-flows, both of which are $1$-forms on $M$. 
V-flows are a covariantisation of the static flows; rather than maximising the flux of a flow $v$ (restricted to the slice $\sigma_{\textrm{sym}}$) through $A$, we maximise the flux of a flow $V$ (which is not restricted to any slice) through $D(A)$. 

\noindent\textbf{Classical V-flow:}
\begin{align}\label{Vflow1}
S(A)=\sup\int_{D(A)}*V& \quad \text{over}\quad (V,\phi) \\
\text{subject to:}
\quad&\phi|_{\mathcal{I}^-}\geq 0\,,\qquad \phi|_{\mathcal{I}^+}\leq1\nonumber\\
&*V|_{\mathcal{I}^+\cup \mathcal{I}^-
}=0\nonumber\\
&\frac{d\phi}{4\GN}\pm V\in j^+\nonumber \\
&d*V=0\,.\nonumber
\end{align}

\noindent\textbf{Classical U-flow:}
\begin{align}\label{Uflow1}
S(A)=\inf\int_{\mathcal{I}^-}*U \quad \text{over}&\quad (U,\psi)\\
\text{subject to:}\quad&\psi|_{D(A)}=1\,,\quad\psi|_{D(A^c)}=0 \nonumber\\
&*U|_{D(A)\cup D(A^c)
}=0\nonumber\\
&U \pm \frac{d\psi}{4\GN}\in j^- \nonumber\\
&d*U=0\,.\nonumber
\end{align}
$j^\pm$ are the sets of future/past-directed causal 1-forms in the cotangent space to a given point in $M$.

V-threads can be thought of as the covariantisation of the static threads of Eq.~\eqref{eq:introRTBIT}, in that they are (typically) spacelike, they connect the boundary region $A$ to its complement (or more precisely connecting their respective causal domains), and their number is maximized subject to a (non-local in time) upper bound on their density.
U-threads are timelike, and their number is \emph{minimized} subject to a (non-local in space) lower bound on their density.

As with all bit thread formulas, the threads have considerable freedom, and there are many optimal thread configurations, but all of them share the feature that the threads are collimated and maximally (for the V-threads) or minimally (for the U-threads) packed at the extremal surface (in this case, the HRT surface).

We refer to the constraints $\frac{d\phi}{4\GN}\pm V\in j^+$ and $U \pm \frac{d\psi}{4\GN}\in j^-$ as the V and U-flows' norm bounds. The scalar functions $\phi$ and $\psi$ enforce the norm bounds.
$\phi$ effectively represents a smeared-out slice, and $\psi$ effectively represents a smeared-out spacetime homology region for $D(A)$. 

The U-flow norm bound requires $U$ to be past-directed causal (and therefore the dual vector field to be future-directed).%
\footnote{Note that our $U$ has the opposite sign from that in~\cite{Headrick:2022nbe}: $U_{\rm here} = - U_{\rm there}$. With our choice of sign, the U-flow lines, the integral curves of the \textit{vector} field, are future-directed.}
The V-flow's norm bound allows it to be spacelike, null or timelike, but optimal V-flows are typically spacelike. There is no counterpart to U-flows in the static prescriptions, but they can be thought of as the timelike counterpart of V-flows. 

Both the V- and U-flow norm bounds have equivalent, non-local forms~\cite{Headrick:2022nbe}. Let $\mathscr{Q}$ be the set of inextendible timelike curves in $M$, and $\mathscr{P}$ the set of spacelike curves $p$ which start on $D(A)$ and end on $D(A^c)$. The V-flow norm bound is equivalent to an upper bound on the integral of the norm of the perpendicular projection of $V$ over inextendible timelike curves:
\bne \forall q \in \mathscr{Q}: \quad \int_q dt |V_\perp| \leq \frac{1}{4\GN}\,, \label{eq:Vnonlocal} \ene
where $t$ is the proper time along $q$  and $V_\perp$ is the projection of $V$ perpendicular to $q$. Compare~\eqref{eq:Vnonlocal} with the classical static norm bound in~\eqref{eq:classs}.
Similarly, the U-flow norm bound is equivalent to the lower bound 
\bne \forall p \in \mathscr{P}: \quad \int_p ds |U_\perp| \geq \frac{1}{4\GN}\,, \label{eq:Unonlocal} \ene
where $s$ is the proper length along $p$ and $U_\perp$ is the projection of $U$ perpendicular to $p$. 

The boundary conditions on $\phi$ in~\eqref{Vflow1} are inequalities and weaker than the equality constraints used in~\cite{Headrick:2022nbe}. The prescriptions are equivalent, essentially because the norm bound encourages optimal $d\phi$ to be as large as possible. The reason we use the inequality forms of the boundary conditions is because they connect more directly to their analogues in the quantum prescriptions.

\section{Quantum covariant bit thread prescriptions} \label{sec:QCBIT}

We now present the quantum covariant thread formulas. In Sec.~\ref{sec:FlowsDerivs}, we will show that they are equivalent to the QES formula, and in Sec.~\ref{sec:OptFlows}, we will describe optimal thread configurations for interesting spacetimes that involve both time-dependence and large quantum effects, such as evaporating black holes.

The change from the covariant classical formulas \eqref{Vflow1}, \eqref{Uflow1} is that, in order to accommodate bulk quantum effects, the divergenceless constraint will be relaxed, just as it was for the static quantum formula.

As in the classical covariant case, we consider smeared-out slices and homology regions, but we do so using a more flexible notion, namely \emph{measures} $\kappa$ on $\mathscr{S}_A$ (for V-flows) and $\rho$ on $\mathscr{R}_A$ (for U-flows). These are probability measures. In other words, they are positive and normalized:
\be
\int_{\mathscr{S}_A}d\kappa(\sigma)=1\,,\qquad
\int_{\mathscr{R}_A}d\rho(r)=1\,.
\ee
These measures can be used to define functions $\phi_\kappa$ and $\psi_\rho$ on $M$ that obey the boundary conditions in \eqref{Vflow1}, \eqref{Uflow1}:
\be
\phi_\kappa :=\int_{\mathscr{S}_A}d\kappa(\sigma)\,\chi_{J^+(\sigma)}\,,\qquad
\psi_\rho :=\int_{\mathscr{R}_A}d\rho(r)\,\chi_r\,.
\label{eq:phipsidefs}
\ee
$J^+(\sigma)$ is the causal future of $\sigma$. For any subset $s\subseteq M$, $\chi_s$ is its characteristic function (equal to 1 on $s$ and 0 elsewhere).
The hypersurfaces of discontinuity for the two characteristic functions in~\eqref{eq:phipsidefs} are the slice $\sigma$ and timesheet $\tau(r)$.

The scalar functions in~\eqref{eq:phipsidefs} are similar but different from those in the classical U- and V-flow prescriptions. They are functions of the measures $\kappa$ and $\rho$. In the classical prescriptions, we smear surfaces using scalar functions, whereas in the quantum prescriptions we smear using measures. When we come to the derivations in Sec.~\ref{sec:FlowsDerivs}, we will see why using measures is necessary.

With this machinery in place, we can now state the two quantum covariant programs:\\ \\
\noindent \textbf{Quantum V-flow:}
\begin{align}\label{Vflow2}
S(A)=\sup\int_{D(A)}&*V \quad \text{over}\quad (V,\kappa)\\
\quad \text{subject to:}\quad&\int_{\sSA}d\kappa(\sigma)=1\nonumber\\
&*V|_{\mathcal{I}^+\cup \mathcal{I}^-}=0\nonumber\\
&\frac{d\phi_\kappa}{4\GN}\pm V\in j^+ \nonumber\\
&\forall r\in\mathscr{R}_A\,,\quad-\int_rd*V\le \int_{\sSA}d\kappa(\sigma)\,S_{\rm b}(r\cap\sigma)\,.\nonumber
\end{align}
The divergence constraint states that the net number of V flow lines ending in any homology region $r$ is upper bounded by the weighted time-averaged entropy (with the weights determined by $\kappa$ or $\phi$). The optimal measures $\kappa$ will only have support on those slices that contain the QES.

As with the static prescriptions, there are loose and strict versions of the quantum V-flow prescription. The prescription in Eq.~\eqref{Vflow2} uses the loose version of the divergence constraint; the \textit{strict} quantum V-flow prescription is identical, except that the divergence constraint is replaced with
\bne \forall r\in\mathscr{R}\,,\quad \left| \int_rd*V \right |\le \int_{\mathscr{S}_A}d\kappa(\sigma)\,S_{\rm b}(r\cap\sigma)\,. 
\ene
This constraint is stricter because: (1) taking the absolute value of the integrated divergence imposes both a lower \textit{and} an upper bound on it, and (2) the constraint is imposed on the larger set of bulk regions $\sR \supset \sR_A$.

By applying the (strict) divergence constraint to $r=M$ and assuming that the bulk state is pure, we get
\bne \int_{M} d*V = 0\,. 
\label{eq:divoc} \ene
This means that there must be as many V-flow sources as sinks in the spacetime. In other words, for every V-flow line that ends at a bulk point, there must be one that appears elsewhere. For loose V-flows, we have $\int_M d*V \geq 0$ instead of~\eqref{eq:divoc}; more V-flow sources are allowed.

We call V-flows that obey the loose divergence constraint \textit{loose flows}, and similarly for \textit{strict flows}. 
\\ \\
\noindent \textbf{Quantum U-flow:}
\begin{align}\label{Uflow2}
S(A)=\inf\int_{\mathcal{I}^-}*U \quad & \text{over}\quad  (U,\rho)\\
\text{subject to:}\quad &\int_{\sR_A}d\rho(r)=1\nonumber\\
&*U|_{D(A)\cup D(A^c)}=0\nonumber\\
&U\pm\frac{d\psi_\rho}{4\GN}\in j^- \nonumber\\
&\forall \sigma\in\mathscr{S}_A\,,\quad-\int_{J^-(\sigma)}d*U\ge \int_{\mathscr{R}_A}d\rho(r)\,S_{\rm b}(r\cap\sigma)\,.\nonumber
\end{align}

The divergence constraint on $U$ states that there is a minimum number of sinks for U-flow lines in the past of every slice $\sigma$, with the lower bound given by the averaged entropy $\int_{\sR_A}\!  d\rho (r) \Sb(r \cap \sigma)$.
The divergence constraint may allow but never requires U-flow sources. This is similar to the loose V-flow prescription's divergence constraint. 

Unlike V-flows, we do not have a pair of loose and strict U-flow prescriptions. Taking inspiration from the strict V-flow constraints, there are two natural but incorrect ways of strengthening the constraints in~\eqref{Uflow2}. Firstly, by taking the absolute value of the integrated divergence, i.e. $-\int d* U \mapsto |\int d* U |$, but this actually weakens the constraint. Secondly, by enlarging the set of homology regions  $\sR_A \mapsto \sR$, but this removes essential dependence on $A$ in the prescription and cannot be QES-equivalent. 

The norm bounds in both quantum flow prescriptions are identical to their classical cousins (except that $\phi_\kappa$ and $\psi_\rho$ are not independent variables; they are dependent on their respective measures). 
The integrated forms of these norm bounds, if they exist (we will not derive them here), must similarly depend on the measures.

The constraints in the quantum V and U-flow prescriptions are not cutoff-independent, because neither $G_N$ nor the bulk entropy $\Sb$ are. We expect that there are covariant cutoff-independent prescriptions, as there are in the static case~\cite{Headrick:2025awv}, but we will not derive them in this paper. We will briefly comment on a promising starting point for such a derivation in Sec.~\ref{sec:OTsupr}.

Note that \eqref{Vflow2} is a concave program and \eqref{Uflow2} is a convex one. Like the pair of classical formulas \eqref{Vflow1} and \eqref{Uflow1}, the quantum V and U-flow prescriptions are Lagrange duals of each other. We leave the proof, which is similar to the one for the classical case \cite{Headrick:2022nbe}, to the reader. 

\section{Optimal flows of quantum covariant bit threads} \label{sec:OptFlows}

In this section, we will discuss optimal flows in the quantum U- and V-flow prescriptions defined in the previous section. An optimal flow is an allowed flow whose boundary flux equals $S(A)$. Studying optimal flows will help us understand how, in practice, the quantum covariant bit thread and QES prescriptions are equivalent.\footnote{In the optimization literature, an optimal value of a variable is often denoted with a star (e.g.\ $x^*$). Following this tradition, optimal V- and U-flows should be denoted $(V^*,\kappa^*)$ and $(U^*,\rho^*)$. To avoid cluttering the formulas, we will not do this; in this section, every flow should be assumed optimal unless it is stated otherwise.}

\subsection{V-flows}
We start with the quantum V-flow prescription given in Eq.~\eqref{Vflow2}. Optimal V-flows obey the constraints and maximise the flux through $D(A)$. 
We will focus on the version of the prescription with the strict divergence constraint, partly because all strict flows are also loose flows, and not vice versa. We will state differences between the prescriptions when they occur.

\subsubsection{General properties of optimal flows} 

\begin{figure}
    \centering
    \includegraphics[width=0.5\linewidth]{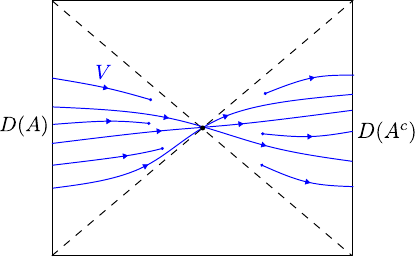}
    \caption{A typical optimal (maximal flux) quantum V-flow.}
    \label{fig:General_Quantum_V_flow}
\end{figure}

The maximal flux through $D(A)$ equals $S(A)$, i.e.\ the generalised entropy of the QES. We will see that the area piece of this generalised entropy is captured by the flux through the QES, and the bulk entropy piece by the net number of flow lines that end in the wedge $\cW(A)$. Fig.~\ref{fig:General_Quantum_V_flow} shows a typical optimal quantum V-flow.

Stokes' theorem tells us that, for any region $r \in \sR_A$, 
\be
\int_{D(A)} *V = -\int_r d*V - \int_{\tau(r)} *V\,.
\ee

The divergence constraint bounds the $\int_r d*V$ term, and the norm bound bounds the $\int_{\tau(r)}*V$ term. So, by applying the constraints, we find that the flux through $D(A)$ is subject to a $\kappa$-dependent upper bound,
\bne \int_{D(A)} *V \leq \inf_{r\in \cR_A} \int_{\sSA} d\kappa (\sigma) \, \Sgen (r\cap \sigma)\,, \label{eq:upper_bound} \ene
which is saturated if and only if the norm and divergence bounds are individually saturated. 

We next show that in any optimal configuration $(V,\kappa)$, $V$ vanishes to the past and future of the QES. Eq.~\eqref{eq:upper_bound} tells us that
\bne 
S(A) = \int_{D(A)} *V \leq \inf_{r\in \cR_A} \int_{\sSA} d\kappa (\sigma) \, \Sgen (r\cap \sigma) \leq  \int_{\sSA} d\kappa (\sigma) \, \Sgen (\cW(A)\cap \sigma) \leq S(A) \,,\label{eq:somearg}
\ene
where $\cW(A)$ is the entanglement wedge. The last inequality uses that the quantum focussing conjecture (QFC) implies that the QES is the $\Sgen$-maximiser on the entanglement horizon $\tau(\cW(A))$. Since the far left and right-hand sides of~\eqref{eq:somearg} are the same, the inequalities in between must be equalities; in particular, $\int_{\sSA} d\kappa(\sigma) \Sgen(\cW(A) \cap \sigma) = S(A)$, which implies that all of $\kappa$'s weight is on slices containing the QES. 
This then implies, by $\phi_{\kappa}$'s definition, that, 
\bne
\phi_{\kappa}|_{I^- (\gamma_{\tsc{QES}})} = 0\, , \quad \phi_{\kappa}|_{I^+ (\gamma_{\tsc{QES}})} = 1\,. \label{eq:phiCond}
\ene
Then, since the norm bound $\frac{d\phi_{\kappa}}{4G_N} \pm V \in j^+$ implies that $V$ must vanish when $\phi_{\kappa}$ is constant, we have that optimal $V$ vanishes in the past and future of the QES:
\bne V|_{I^{\pm}(\gamma_{\tsc{QES}})} = 0\,.\label{eq:Vopc} \ene
This is a property that optimal quantum V-flows (loose or strict) share with optimal classical V-flows~\cite{Headrick:2022nbe}. The reason that they share this property is that their norm bounds are the same.

Next, for~\eqref{eq:upper_bound} to be saturated for an optimal measure $\kappa$, the divergence constraint (loose or strict) must be saturated for any homology region whose timesheet contains the QES. 
Taking $r = \cW(A)$, we get that the net number of threads ending in $\cW(A)$ equals the bulk entropy:
\bne \int_{\mathcal{W}(A)}\!\!\!\!\! d *V + S_b (\cW(A) \cap \sigma) = 0 \label{eq:ofsat}\,, \ene
where $\sigma$ is any one of the optimal slices that contains the QES.

Also, the flux of $V$ through the QES equals $|\gammaQES|/4G_N$, because the norm bound is saturated. Together, these account for the area and bulk entropy pieces of the generalised entropy of the QES. 

Where the threads can start and end is controlled by the entanglement structure of the bulk state, as we will explore next. 

\subsubsection{Bell pair example} \label{sec:VBell}

The bulk state (in particular, its subregion entropies) determines where quantum V-flow lines can start and end. Let us explore this in a simple toy example.

Consider a typical bulk state and spacetime, and add a Bell pair of particles created in the past of the QES, with one of the pair passing through the entanglement wedge $\mathcal{W}(A)$, and the other through $\mathcal{W}(A^c)$. For the purposes of this discussion, we will neglect the small shift in the QES's position from the Bell pair creation (we will study an example where the shift is not small in Sec.~\ref{sec:V_Islands}), and we will treat the particles as following classical trajectories.

\begin{figure}
    \centering
    \begin{overpic}[width=0.6\linewidth]{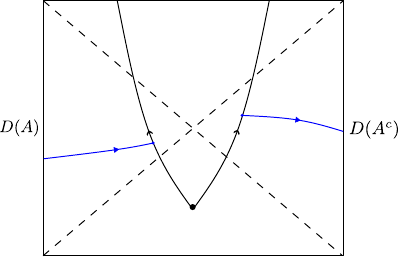}
    \end{overpic}
    \caption{The flow lines (blue curves) of the additional maximal V-flux when a Bell pair of particles (the black timelike bulk curves) is created in the past of the QES. The flow lines have to end on the segment of the particle worldline in $\cW(A)$, and to reappear on the worldline in $\cW(A^c)$, because of the divergence constraint. }
\label{fig:EPR_pair_V_flow}
\end{figure}

In Fig.~\ref{fig:EPR_pair_V_flow}, we depict the extra V-flow lines allowed for by adding the Bell pair to the bulk state.\footnote{In these and other diagrams in this section, we show only the simplest representatives amongst the set of all optimal flows. These ``simple" flows do not have closed V-flow line loops (which contribute nothing to the flux through $D(A)$ and only make it harder to satisfy the norm bound), or more than the minimal number of sources and sinks.}
The Bell pair increases the entanglement across the QES, and allows for additional flux through $D(A)$, because the divergence constraint allows for additional sinks in the wedge $\cW(A)$. 
These extra flow lines cannot enter, start or end in $J_{\pm}(\gamma_{\tsc{QES}})$, because of~\eqref{eq:Vopc}; both the original and extra V-flow lines are constrained to be in the union $\cW(A) \cup \cW(A^c)$ of entanglement wedges. 

The extra flow lines effectively jump across the QES. Where they are allowed to jump to and from depends on the Bell particles' worldline trajectories. 
First note that, to increase the flux through $D(A)$ without violating the norm bound on the QES (since the norm bound is already saturated there), the extra flow lines have to end in $\cW(A)$. Moreover, the flow lines must end on the worldline of the particle that enters $\cW(A)$, at any point on the worldline, because if it ended anywhere else in $\cW(A)$, then we would have a violation of the divergence bound, as there are many homology regions $r$ that include that point but not the entangled particle. Having ended on the worldlines in $\cW(A)$, the flow lines must reappear on the worldline in $\cW(A^c)$, because the bulk entropy of the intersection between any slice and the interior of a pair of tubes containing the Bell pair is zero. 

Both the loose and strict divergence constraints require sources on the worldline in $\cW(A^c)$ for optimal V-flows. The loose constraint allows for more than the minimum number of sources in $\cW(A^c)$ for optimal flows, but these will not affect the flux through $D(A)$, since these flow lines must end on $D(A^c)$.

There is another way in which the optimal V-flow depends on the creation point and trajectory of the Bell pair. In contrast to the case shown in Fig.~\ref{fig:EPR_pair_V_flow}, no additional flux would be allowed through $D(A)$ if both particles passed through the same entanglement wedge, or if the entangled pair of particles were created in any of the other causal regions ($\cW(A)$, $\cW(A^c)$, or $J^+(\gamma_{\tsc{QES}})$). In the case that both particles passed through or were created in $\cW(A)$, while this would allow flow lines to jump from one particle worldline to another, it would not increase the flux through $D(A)$, since these extra flow lines have nowhere to go but return to $D(A)$, because of the saturated norm bound on the QES.

\subsubsection{Islands} \label{sec:V_Islands} 
Islands are connected components of an entanglement wedge that are disjoint from the AdS boundary~\cite{Almheiri:2019psf,Penington:2019npb,Almheiri:2019qdq}. 
In the static quantum prescriptions, islands manifest as bulk regions where so many quantum bit threads reappear that the region's boundary is a bottleneck for threads leaving the region~\cite{Rolph:2021hgz,Headrick:2025awv}. This is an interesting manifestation of islands, and we will see that it continues to hold for covariant quantum bit threads.

\begin{figure}
    \centering
    \begin{overpic}[width=0.7\linewidth]{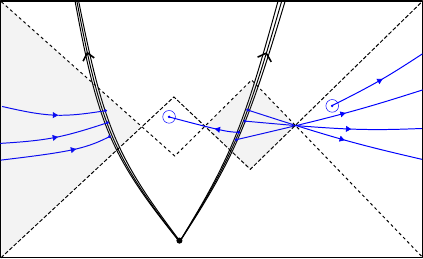}
    \put(5,17){$D(a)$}
    \put(56.5,25.7){\small $D(I)$}
    \put(-10,30){$D(A)$}
    \put(101,30){$D(A^c)$}
    \end{overpic}
    \caption{An optimal V-flow in an island phase. We create an island by adding order $G_N^{-1}$ spatially coincident Bell pairs (the black timelike curves). The shaded bulk region is the entanglement wedge $\cW(A) = D(a) \cup D(I)$. 
    The blue curves are the ``quantum" part of the optimal V-flow configuration: the subset of flow lines that jump across the QES and capture the bulk entropy contribution to $S(A)$. 
    The $\odot$ denote points where a thread has gone into/out of the page; this thread is continuous and reaches the target $D(A^c)$ by going around the island $D(I)$. 
    }
    \label{fig:Vislands}
\end{figure}

We engineer a setup with an entanglement island by building on the single Bell pair example of Sec.~\ref{sec:VBell}. We simply keep adding more Bell pairs (following identical trajectories) until the number is sufficiently high, $O(G_N^{-1})$, that the generalised entropy is lowered by including in the wedge $\cW(A)$ a boundary-disjoint connected component $D(I)$ containing the Bell pair partners; this region is the island.
This setup, and the resulting optimal V-flow, are depicted in Fig.~\ref{fig:Vislands}.

To avoid superfluous details in the figure, we only show the additional V-threads allowed for by the addition of the Bell pairs (neglecting backreaction and positional shifts of $\del D(a)$). 
There are two types of flow line endpoints in the figure: (1) where a thread has gone into/out of the page, and (2) a true sink and source, as allowed for by the divergence constraint. 
The $\odot$ symbols distinguish the former from the latter. 

The divergence constraint allows $S_b (a)$ of the V-flux from $D(A)$ to end in $D(a)$, but the flow lines are forced to reappear in the bulk region that is entangled with $D(a)$, because of the divergence constraint. When so many flow lines reappear in the bulk region that its boundary is a new geometric bottleneck (because of the norm bound), then that bulk region is an island, and the new bottleneck is the boundary of the island. This is how islands manifest in the quantum V-flow prescription: speaking informally, an island is a highly entangled bulk region which bit threads jump into and get ``stuck" trying to exit.

\subsubsection{Evaporating black holes} \label{sec:V_BHs}

\begin{figure}[htbp]
	\centering
	\begin{subfigure}[b]{0.6\textwidth}
		\centering     
		\includegraphics[width=\textwidth]{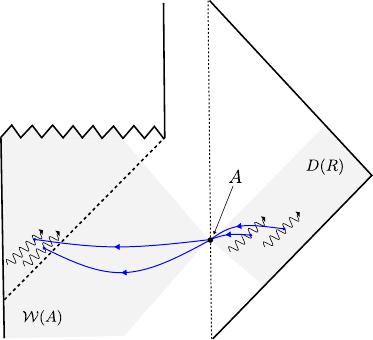}
		\caption{Before the Page time.}
		\label{fig:VbhiA}
	\end{subfigure}
	\hfill
	\begin{subfigure}[b]{0.6\textwidth}
		\centering  
		\includegraphics[width=\textwidth]{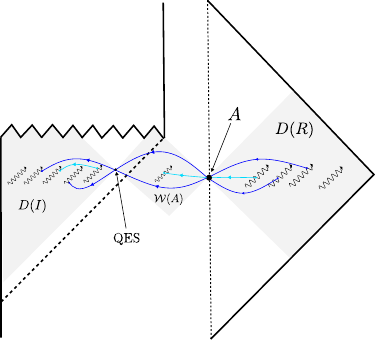}
		\caption{After the Page time.}
		\label{fig:VbhiB}
	\end{subfigure}
	\caption{Optimal V-flows for an evaporating AdS black hole coupled to a flat space bath. 
		The V-flow lines have sinks on the worldlines of those Hawking quanta in the black hole interior that are entangled with quanta in the radiation region $D(R)$ (on which the strict version of the divergence constraint also requires there to be sources).
	}
	\label{fig:Vbhi}
\end{figure}

We can take what we have learned so far about optimal quantum V-flows and apply it to evaporating black holes in AdS spacetimes. These spacetimes are not time-reflection symmetric, which is why we need our covariant prescriptions to study them. In Fig.~\ref{fig:Vbhi}, we illustrate what we will explain in words next: our evaporating black hole spacetime and typical maximal V-flows both before and after the Page time. 

The spacetime setup is an extension of that described in Sec.~\ref{sec:STSTP}. 
The full spacetime $M$ has both an asymptotically AdS component and an appended ``bath" component: $M = M_{\rm AdS} \cup M_{\rm bath}$.
We identify the AdS conformal boundary $N$ of $M_{\textrm{AdS}}$ with the timelike component of $\del M_{\rm bath}$.
This spacetime corresponds to a holographic setup where the non-gravitational description is a field theory on $M_{\rm bath}$ with holographic degrees of freedom on $\del M_{\rm bath}$ such that the dual is gravity in an asymptotically AdS spacetime $M_{\textrm{AdS}}$ coupled to a non-gravitational theory on $M_{\rm bath}$. This setup reduces to the standard setup of Sec.~\ref{sec:STSTP} if we take $M_{\rm bath}$ to be empty.

A consequence of extending $M$ with a bath region is that subregion $A$ of the AdS boundary $N$ and its domain of dependence $D(A;M_{\rm bath})$ within the non-holographic spacetime are the same: $A = D(A;M_{\rm bath})$ (and neither equal $D(A;N)$).
$D(A;M_{\rm bath})$ is the relevant domain of dependence for holographic entanglement entropy, because time slices of $D(A;N)$ do not have the same entropy with these open AdS boundary conditions. From this point on, in this open boundary condition setup, we use $D(A)$ to denote $D(A;M_{\rm bath})$ ($=A$). The important consequence for the V-flow prescription is that, to calculate $S(A)$, we maximise the flux through $D(A) = A$ (and not $D(A;N)$) into $M_{\rm AdS}$. 

There is a choice as to whether to extend the domain of the V and U vector fields beyond $M_{\rm AdS}$ to include the bath region. The simpler choice is not to; then one can effectively ignore the bath region and treat the setup as an asymptotically AdS spacetime with open boundary conditions, and all that changes in the formulas is to replace $D(A)$ by $A$. In contrast, extending the flow prescriptions to include the bath requires modifying the definitions in Sec.~\ref{sec:STSTP}, because, for example, the AdS conformal boundary $N$ is now an internal boundary in the full spacetime $M$. The modifications, such as how the slices and timesheets extend into the bath, are straightforward and natural. Because $\GN =0$ in the bath, there is no norm bound on the V-flow there. In contrast, the norm bound on $U$ becomes infinitely strong: either $d\psi_\rho = 0$ or $U$ diverges. Extending the flows into the bath is appealing on a conceptual level because it allows threads to connect Hawking quanta in the black hole interior with their entangled partners in the bath (see Fig.~\ref{fig:Vbhi}).

In an optimal configuration, there cannot be any V-flux on the AdS boundary anywhere except through $A$. In particular, flow lines coming into $M_{\rm AdS}$ from $A$ cannot circle back and pass back through $N\backslash A$ elsewhere. 
This is because the norm bound requires that $V = 0$ wherever $\phi_\kappa$ is constant, so $V = 0$ at any point in $M$ that is not contained in any $\sigma \in \sSA$.

As always, where the flow lines are allowed to start and end depends on the subregion entropies of the bulk state through the divergence constraint. All V-threads from $A$ must end on sinks; there is nowhere else for them to end.

We can consider a one-parameter family of boundary regions $A(t)$ given by the intersection of slices in the time foliation $M = \bigcup_t \Sigma_t$ with the AdS boundary $N$. 
As a function of boundary time $t$, the maximum V flux through $A(t)$ will give the Page curve.
The radiation region $R := \Sigma_t \backslash A$ is the complement region on the Cauchy slice in $M_{\rm bath}$, and
the entanglement wedge of the radiation region includes an entanglement island after the Page time: $\cW(R)= D(R) \cup D(I)$.
There is a transition in the set of optimal V-flows at the Page time:
 
\begin{enumerate}
    \item Before the Page time, there are relatively few entangled Hawking quanta shared between the black hole interior and the radiation region $D(R)$. The QES is the empty set, and the boundary entropy approximately equals the entropy of the Hawking modes in the black hole interior, $S(A) \approx \Sb(\textrm{BH interior})$. As depicted in Fig.~\ref{fig:VbhiA}, the divergence constraint allows the V-flow lines to have sinks on those interior Hawking quanta that are entangled with quanta that have reached $D(R)$ by time $t$ (on which the threads must reappear for the strict constraint), and this is the mechanism by which the maximum V-flux through $D(A)$ equals the entropy of the interior Hawking modes.
    
    \item After the Page time, there are $O(\GN^{-1})$ entangled Hawking quanta shared between the black hole interior and $D(R)$. The divergence constraint allows for many V-flow sinks in the interior, because of its large entropy. Each Hawking quantum in the black hole interior that is entangled with a partner in $D(R)$ is a potential V-flow sink. However, as depicted in Fig.~\ref{fig:VbhiB}, the norm bound causes there to be a geometric bottleneck for the flow, such that not all of the sinks can be ``used". This bottleneck is disjoint from the AdS boundary, the norm bound is saturated on it, and it is the (non-empty) QES. The fact that the norm bound does not allow all interior modes to have threads connecting them to $A$, and because the max flux through $A$ is limited by the size of the bottleneck, is why the max V-flux (approximately) equals the area of the bottleneck, 
    \bne \max \int_{D(A)} *V \approx \frac{\rm{Area}(\gammaQES)}{4G_N}\,,\ene
    rather than the bulk entropy $\Sb (\textrm{BH interior})$, which is larger.

    The V-threads also capture the contribution to $S(A)$ from the bulk entropy $\Sb (a)$.
    Suppose we have Hawking quanta in the black hole interior whose exterior partners are in the wedge $\cW(A)$. Those exterior quanta have had insufficient time to reach $D(R)$, and they cause there to be 
    non-zero bulk entropy $\Sb(a)$. Then, the divergence constraint allows threads connected to $A$ to end on these entangled quanta in $\cW(A)$. These threads effectively jump across the QES and into the black hole interior, because they must reappear on the Hawking quanta with which the quanta in $\cW(A)$ are entangled. Finally, they can leave $M_{\rm AdS}$ by ending on quanta in the black hole interior which are entangled with quanta in $D(R)$ (i.e.\ the old radiation). 
    A flow line representing this process is depicted in Fig.~\ref{fig:VbhiB} with a light blue curve.
    For the divergence constraint to be satisfied, all threads that end on quanta in $M_{\rm AdS}$ must reappear on their entangled partner quanta in $D(R)$.%
    \footnote{There are always enough old interior modes for the threads to terminate on. If there were not, this would imply that $\Sb (\text{old}) < \Sgen(a)$, which is inconsistent with the QES being non-empty, which requires that $\Sgen (a) < \Sb (a \cup \text{new} \cup \text{old} ) ( \approx \Sb(\text{old}))$.}
    The loose constraint allows for more V-flow sources than the strict constraint, but this does not increase the max-flux through $A$.
    
\end{enumerate}

We have approximated the bulk state on post-black hole formation time slices as a collection of Bell pair excitations, with definite particle numbers of interior and exterior modes, to simplify this discussion. This is a good approximation for weakly coupled bulk fields, though a slightly more accurate one would be the two-mode squeezed state
(which has non-zero variance in particle number), plus corrections from interactions. 
The differences between these bulk state approximations are unimportant, to an extent, because the max V-flow is insensitive to certain details of the bulk state; the boundary entropy $S(A)$ is invariant under unitary conjugation of the bulk state reduced to an optimal slice of the wedge $\cW(A)$. 
That said, the set of allowed V-flows that achieve that max V-flux is \textit{not} invariant under such conjugations, because the divergence constraint is sensitive to the entropies of more than just the wedge $\cW(A)$. 

This connection --- how the bulk subregion entropies are encoded in the set of allowed flows and vice versa --- was explored in detail and generality for the static quantum prescriptions in~\cite{Headrick:2025awv}. We leave the extension of this analysis to our covariant prescriptions for future work.

\subsection{U-flows}

\subsubsection{General properties of optimal flows}
Now we turn to optimal U-flows. The quantum U-flow prescription is~\eqref{Uflow2}.

As the norm bound is saturated on the QES for optimal flows, we have 
\bne \psi_{\rho}|_{\mathcal{W}(A)} = 1\, , \quad \psi_{\rho}|_{\mathcal{W}(A^c)} =0\,. \ene
So the measure $\rho$ is constrained to have support only on timesheets containing the QES,
\bne \rho (r) > 0 \implies \gammaQES \subset \tau(r)\,, \label{eq:tauconstr} \ene
and optimal flows vanish in the entanglement wedges:
\bne U |_{\mathcal{W}(A)} = U |_{\mathcal{W}(A^c)} = 0\,. \ene

The U-flow's divergence constraint holds for all slices $\sigma$. For optimal flows, when we apply the divergence constraint to the earliest slice containing the QES, the constraint is saturated.
This means that the divergence constraint forces there to be U-flow line sinks in the past of the QES, with the net number of sinks equal to the entropy across the QES. Because U-flow lines are timelike, flow lines which end in $J^- (\gamma_{\tsc{QES}})$ do not pass through the QES, and either originate from $\cI^-$ or from a source (which must be matched with another sink, to satisfy the divergence constraint). The quantum U-flow divergence constraint increases the minimal U flux through $\cI^-$ (in comparison to the classical constraints). This additional flux ends in $J^- (\gamma_{\tsc{QES}})$ and so cannot help with satisfying the norm bound on the QES. This additional ``quantum" contribution to the minimal boundary flux is how the quantum U-flow prescription captures the correction to $S(A)$ from the bulk entropy.

\paragraph{$\mathbf{U}$-flows and UV-divergent entropies.} 
The divergence constraint requires a UV-divergent number of U-flow sinks for finite-energy bulk states, because their subregion entropies have universal area-law divergences. For a $D$-dimensional bulk,
\bne \Sb (\sigma \cap r) \approx \frac{\textrm{Area}(\sigma \cap \tau(r))}{\epsilonUV^{D-2}}. \label{eq:Sbvac} \ene
First consider the earliest slice $\sigma$ in $\sSA$, denoted $\sigma_-$, which is $\del I^- (\del A)$ for closed AdS boundary conditions, and the past boundary of the Wheeler-DeWitt patch of $A$ for the black hole setup. 
This is an important slice for the divergence constraint, because sinks in the past of this slice are in the past of all slices in $\sSA$.
While there must be at least net $\int_{\cR_A} d\rho(r) S_{b} (r \cap \sigma_-)$ sinks in $J^-(\sigma_-)$, the position of these sinks within $J^- (\sigma_-)$ is not constrained. When $J^-(\sigma_-) = \sigma_- = \tilde{\mathcal{I}}^-$, such as when $A$ is a boundary Cauchy slice of one of the AdS boundaries in an eternal two-sided AdS black hole, then there must be a UV-divergent number of sinks on $\tilde{\mathcal{I}}^-$. 

As a heuristic, one can think of there being three components of optimal U-flows: 
\begin{enumerate}
	\item The ``classical" component, whose flow lines are continuous and all pass through the QES, saturating the norm bound there. Its $O(\GN^{-1})$ flux through $\cI^{-}$ is typically the largest component of the flux, and captures the area term in $\Sgen$.
	\item The ``quantum" component, whose flow lines are discontinuous and may effectively jump from the past to the future of the QES, and whose flux through $\cI^-$ captures the bulk entropy piece of $\Sgen$. The quantum component can be further subdivided:
	\begin{enumerate}
		\item A UV-divergent number of flow lines, required because the divergence constraint, together with the UV-divergent terms in the bulk entropies, necessitates an $O(\epsilon_{\tsc{UV}}^{-(D-2)})$ number of sinks. Let us call this the vacuum quantum U-flow component, since the origin of these UV divergences is the short-distance entanglement structure of vacuum states.
		\item An additional finite number of flow lines allowed for by the finite vacuum-subtracted piece of the bulk entropies.  
	\end{enumerate}
\end{enumerate}	

Not all of the UV-divergent number of sinks of the vacuum quantum U-flow component have to be in $J^-(\sigma)$ of the earliest slice in $\sSA$. 
This is because, for a fixed timesheet $\tau$ that contains the QES, the area piece of the bulk entropy, Eq.~\eqref{eq:Sbvac}, is approximately maximised for slices $\sigma$ which contain the QES, as follows from classical maximin (assuming, to simplify this discussion, that the QES and HRT surfaces are close to each other; that the bulk entropy gradients are not too large). The larger bulk entropy $\Sb(\sigma \cap r)$ of these later time slices allows the U-threads to have their vacuum entropy sinks to the future of the earliest slice.

Note that optimal V-flows also have a UV-divergent number of sinks, to capture that piece of the bulk entropy, but these sinks can be close to the QES. In contrast, the U-flow sinks are far away from the QES (to the past). 

\subsubsection{Bell pair example} \label{sec:U_Bell_Pair}

\begin{figure}
    \centering
    \begin{overpic}[width=0.5\linewidth]{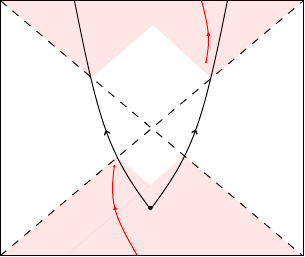}
    \put(32,14){\Large \textcolor{red}{$U$}}
    \put(7,2){ $J^-(p_1) \cup J^-(p_2)$}
    \put(33,79){$J^+(p_3) \cup J^+(p_4)$}
    \put(34.5,32.5){\small $p_1$}
    \put(62,33){\small $p_2$}
    \put(26,57){\small $p_3$}
    \put(70,57){\small $p_4$}
    \put(-11,40){$D(A)$}
    \put(101,40){$D(A^c)$}
    \end{overpic}
    \caption{The flow lines (red curves) of the additional minimal U-flux when a Bell pair of particles is created in the past of the QES. 
    The region $Y_{\textrm{sinks}}(\ln 2) = J^-(p_1) \cup J^-(p_2)$ is where the additional U-flow must have $\ln 2$ sinks, and $Y_{\textrm{sources}}(\ln 2) = J^+(p_3) \cup J^+(p_4)$ is where it can have $\ln 2$ sources.}
    \label{fig:EPR_pair_U_flow}
\end{figure}

Now we revisit the single Bell pair example, this time for the U-flow prescription. In Fig.~\ref{fig:EPR_pair_U_flow}, we illustrate the extra flux through $\cI^-$ required by the additional entanglement from the Bell pair.
The Bell pair increases the minimal U-flux through $\cI^-$ by $\ln 2$, because the divergence constraint requires additional sinks. 

Determining where optimal U-flows have sources and sinks is in some ways more difficult than for V-flows. We will describe the general case, then give the Bell pair setup as a special case. 
Define the set of slices
\bne \mathscr{E}_A(r,s) := \{ \sigma \in \sSA | \Sb (\sigma \cap r) \geq s \}\, . \ene
The divergence constraint requires at least net $s$ U-flow sinks in $J^-(\sigma)$ for every $\sigma \in \mathscr{E}_A (r,s)$, for a fixed measure $\rho$ with all its weight on a single $r$. 
Next we define the spacetime region
\bne Y_{\rm{sinks}}(s) := \bigcup_{r \in \cR_A^*} \bigcap_{\sigma \in \mathscr{E}_A(r,s)} J^{-} (\sigma)\,. \label{eq:Ysind}\ene
This is the region where optimal U-flows must have at least net $s$ sinks. $\cR_A^*$ is the set of optimal homology regions (whose timesheets necessarily contain the QES).
The union over $r \in \cR_A^*$ in the definition~\eqref{eq:Ysind} originates from the U-flow prescription requiring only the existence of an optimal $\rho^*$, a subset of which have weight on a single $r\in \cR_A^*$. 

We can also define the spacetime region where there can be up to net $s$ sources:
\bne Y_{\rm{sources}}(s) := \bigcup_{r \in \sR_A^*} \bigcap_{\sigma \in \mathscr{E}_A(r,s)} J^{+} (\sigma)\,. \ene
For these two definitions, note that, mathematically, we could have used the smallest subsets $\mathscr{E}_A^\pm \subseteq \mathscr{E}_A$ for which 
\bne \bigcap_{\sigma \in \mathscr{E}^\pm_A(r,s)} J^{\pm} (\sigma) = \bigcap_{\sigma \in \mathscr{E}_A(r,s)} J^{\pm} (\sigma) \,.\ene
This reflects the fact that a U-flow sink to the past of a slice helps satisfy the divergence constraint for all slices to the future of that slice.

The sink and source regions are relatively simple for our Bell pair state. For this state,
\bne Y_{\rm{sinks}}(s) = \begin{cases} M & s\leq 0 \\
J^-(p_1) \cup J^-(p_2) & 0< s \leq \ln 2 \\
\varnothing & \ln 2 < s \end{cases} \label{eq:BellSink}\ene
and 
\bne Y_{\rm{sources}}(s) = \begin{cases} M & s\leq 0 \\
J^+(p_3) \cup J^+(p_4) & 0< s \leq \ln 2 \\
\varnothing & \ln 2 < s \end{cases} \,.\label{eq:BellSource} \ene
$p_1$ and $p_2$ are the points where the Bell pair enter the entanglement wedges $\cW(A)$ and $\cW(A^c)$, and $p_3$ and $p_4$ are where they leave them. $\ln 2$ is the maximum bulk entropy $\Sb(\sigma \cap r)$.

Let us explain Eq.~\eqref{eq:BellSink} more fully. In the definition~\eqref{eq:Ysind}, we take a union over optimal homology regions (those whose timesheets contain the QES). There is the optimal homology region $r_1^* = \cW(A)$, for which
the earliest slice $\sigma$ for which $\Sb (\sigma \cap r) = \ln 2$ is the past light cone $\del J^-(p_1)$ of the point $p_1$ where the left particle first intersects $\cW(A)$. There is also the optimal homology region $r_2^* = (\cW(A^c))^c$, giving the past light cone $\del J^- (p_2)$ for the right particle. In determining the sink region $Y_{\textrm{sinks}}(s)$ for $0<s\leq\ln 2$, we need not consider any other $r^*$; the sink region is not enlarged by considering any other optimal homology region whose timesheet $\tau$ is in between these two extremes, because such timesheets cannot intersect either of the two particles to the future of the two past light cones. 

The essential reason why none of the extra $\ln 2$ U-flow lines can end in the future of $Y_{\rm{sinks}}(\ln 2)$ is because then there would exist a slice whose past includes the particle creation point but not all the sinks, which violates the divergence constraint. This is similar to why the earliest that sources can appear is the past boundary of $Y_{\rm{sources}}(\ln 2)$.

Where the Bell pair enters the entanglement wedges, rather than where the pair is created, is, in a sense, more relevant for where the U-flow line can end. For example, the extra U-flow must end in the past of $p_1$ or $p_2$, $J^-(p_1) \cup J^-(p_2)$, and this region includes not just the past of the pair creation point $J^- (p_{\rm creat.})$, but also a portion of $J^+(p_{\rm creat.})$ (for generic timelike particle worldlines).

There is a special subset of optimal U-flows, which are ``inextendible". Consider the set $\mathscr{U}_{\rm div}$ of U-flows which obey the divergence constraint, but not necessarily the norm bound, and $\mathscr{U}$, the set of U-flows which obey both constraints. We can argue that, in  $\mathscr{U}_{\rm div}$, there is guaranteed to be at least one ``inextendible" U-flow that \textit{does} obey the norm bound, and is therefore in $\mathscr{U}$. 
By inextendible, we mean a $U \in \mathscr{U}_{\rm div}$  whose flow lines cannot be extended, either to sources further in the past or to sinks further in the future, without violating the divergence constraint.%
\footnote{
In our Bell pair example, the latest the additional U-flow line can end is either $p_1$ or $p_2$, and the earliest it can reappear is either $p_3$ or $p_4$. The flow lines of inextendible U-flows have sinks where entanglement between $\cW(A)$ and $\cW(A^c)$ is created, in a sense, and vice versa for when the Bell pair leave the entanglement wedges.
}
These inextendible U-flows have the pointwise largest norms $|U|$'s amongst $\mathscr{U}$, and are therefore, in a sense, most compatible with the norm bound. Any extension of a $U \in \mathscr{U}$ is also in $\mathscr{U}$. A corollary of this is that, in the set of inextendible U-flows, there is guaranteed to be at least one in $\mathscr{U}$, because otherwise $\mathscr{U} = \varnothing$, which is inconsistent with our proof of QES-equivalence.

\subsubsection{Islands}\label{sec:U_islands}

\begin{figure}
    \centering
    \begin{overpic}[width=0.8\linewidth]{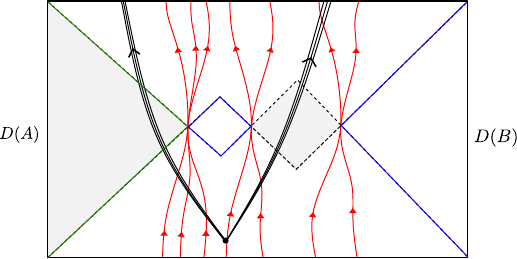}
    \put(45,29){\textcolor{blue}{$\tau_2$}}
    \put(72,35){\textcolor{blue}{$\tau_2$}}
    \put(18,42){\textcolor{OliveGreen}{$\tau_1$}}
    \put(18,24){$D(a)$}
    \put(57,24){\small $D(I)$}
    \put(91,23){\colorbox{white}{$D(A^c)$}}
    \end{overpic}
    \caption{An optimal U-flow when there is an entanglement island.}
    \label{fig:U_flow_island}
\end{figure}

We now turn to how entanglement islands manifest in the quantum U-flow prescription. In Fig.~\ref{fig:U_flow_island}, we show an optimal U-flow configuration in the same island setup as we used for V-flows in Sec.~\ref{sec:V_Islands}. 

When the bulk state has few Bell pairs, there is no island. This situation is similar to that of the single Bell pair example in Sec.~\ref{sec:U_Bell_Pair}, except now with $k \ln (2)$ additional flux through $\cI^-$ ending at sinks before reaching the QES, where $k$ is the number of Bell pairs. 
For small $k$, an optimal homology region is $r= D(a)$ (see Fig.~\ref{fig:U_flow_island}).  Note that the optimal homology region is generally non-unique. 

As we add more Bell pairs to the bulk state, the allowed U-flows have more flow lines attached to $\cI^-$ ending on sinks to the future, because the divergence constraint requires additional sinks as the entropy increases. 

We induce an island phase transition by adding a sufficiently large number of Bell pairs, $k \gtrsim \GN^{-1}$. There is a corresponding phase transition in the set of optimal U-flows. Consider two candidate homology regions, $r_1 = D(a)$ and $r_2$, whose timesheets are $\tau_1$ and $\tau_2$ (see Fig.~\ref{fig:U_flow_island}). Relative to $r_1$, we can reduce the minimum flux through $\cI^-$ by switching to $r_2$. This is because the entropy $\Sb (\sigma \cap r_2)$ is zero for all slices $\sigma$. So, while choosing $r_2$ requires many more ``classical" flow lines (which capture the island boundary area), this is outweighed by the reduction in the number of U-flow sinks required by the divergence constraint.

The set of optimal U-flows changes discontinuously at the island phase transition as the bulk entanglement increases. The transition is from ``quantum" U-threads with sinks accounting for the bulk entropy $\Sb(a)$, to divergenceless ``classical" U-threads accounting for the island area $|\del I|/4\GN$.%
\footnote{A discontinuous transition also happens for \emph{classical} U-flows, whenever the HRT surface changes discontinuously as a function of, for example, the size of the boundary region $A$.} 
This is different from the set of optimal quantum V-flows, whose character changes continuously at the island phase transition.
This discontinuous change in the set of optimal U-flows is how the island phase transition manifests in the quantum U-flow prescription.

\subsubsection{Evaporating black holes}

Now we consider optimal quantum U-flows in the same evaporating black hole setup where we analysed optimal V-flows in Sec.~\ref{sec:V_BHs}. 

\begin{figure}[htbp]
     \centering
     \begin{subfigure}[b]{0.45\textwidth}
         \centering        
         \includegraphics[width=\textwidth]{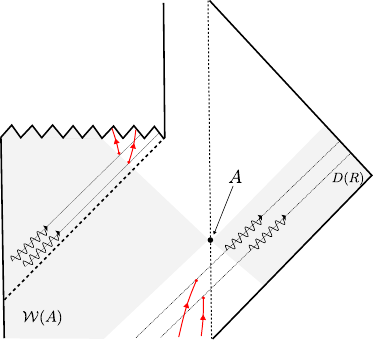}
         \caption{Before the Page time. 
         }
         \label{fig:UbhiA}
     \end{subfigure}
     \hfill
     \begin{subfigure}[b]{0.45\textwidth}
         \centering        
        \includegraphics[width=\textwidth]{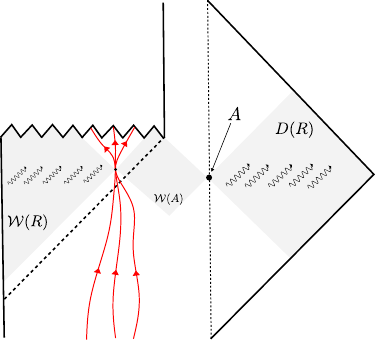}
         \caption{After the Page time.}
         \label{fig:UbhiB}
     \end{subfigure}
    \begin{subfigure}[b]{0.45\textwidth}
         \centering    
        \includegraphics[width=\textwidth]{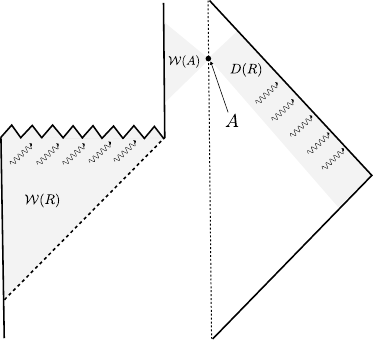}
         \caption{Post-evaporation.}
         \label{fig:UbhiC}
     \end{subfigure}
        \caption{
        Typical optimal U-flows for an evaporating AdS black hole coupled to a flat space bath. Before the Page time, the U-threads end before entering the entanglement wedge $\cW(A)$, and ``jump" forward in time. After the Page time, the threads are continuous and pass through the QES. 
        }
    \label{fig:Ubhi}
\end{figure}

In Fig.~\ref{fig:Ubhi}, we show typical optimal U-flows both before and after the Page time. 
Let us discuss the key features:
\begin{enumerate}
    \item Before the Page time, the situation is similar to that considered in the single Bell pair example (Sec.~\ref{sec:U_Bell_Pair}). The U-flow's divergence constraint requires there to be sinks for the flow in the past of where entanglement entropy increases. The U-flux through $\mathcal{I^-}$ grows with increasing boundary time $t$, because there are more entangled Hawking quanta in $D(R)$, which increases the bulk entropy 
    and forces there to be more sinks for the U-flow lines. 
    The QES is the empty set and, equivalently, nowhere that the U-flow's norm bound needs to be saturated.
    \item After the Page time, the situation is similar to the island setup of Sec.~\ref{sec:U_islands}. The U-flux through $\cI^-$ is small when the measure $\rho$ has support on a homology region with small entropies $\Sb(\sigma \cap r)$ for any $\sigma \in \sSA$, such as $r = \cW(A) $ (see Fig.~\ref{fig:UbhiB}).
    Homology regions like $r = M_{\rm AdS}$ are not optimal because they have large entropies due to the accumulation of Hawking quanta in $D(R)$, and so require many U-flow sinks. 
    At the Page time, the set of optimal U-flows changes abruptly, from ``quantum" flows that ``jump" in time across the Wheeler-DeWitt patch, to divergenceless ``classical" flows that pass through the QES,%
    \footnote{The continuous threads capture the bulk area term of $S(A)$ and dominate the flux through $\cI^-$, but there are also generically a number of discontinuous U-threads attached to $\cI^-$ which account for the bulk entropy term.}
    similar to the island phase transition discussed in Sec.~\ref{sec:U_islands}.
\end{enumerate}

The black hole has fully evaporated from the boundary perspective for Cauchy slices on the AdS boundary $N$ which are strictly to the future of the black hole evaporation endpoint (see Fig.~\ref{fig:UbhiC}). At these late times, the boundary entropy is $S(A) \approx 0$. From the perspective of the QES prescription, $S(A) \approx 0$ because the QES is empty and the bulk time slice contains only those few quanta which have not yet escaped into the bath. For consistency, we should also get that the minimal U-flux through $\cI^-$ is approximately zero. 
The wedge $\cW(A)$ in Fig.~\ref{fig:UbhiC} is an optimal homology region for the U-flow prescription, because $\Sb (\cW(A)\cap \sigma) \approx 0$ for all $\sigma \in \sSA$, so U-flow sinks are not required by the divergence constraint. $U \approx 0$ is an allowed and therefore optimal U-flow.

\section{Derivations of the quantum covariant bit thread prescriptions} \label{sec:FlowsDerivs}

\subsection{Concave-convex quantum maximin}
\label{QuantumMaximin}

The purpose of this subsection is to reach the starting point for our derivation of the quantum covariant bit thread prescriptions: a QES-equivalent, surface-based prescription that is concave-convex in its variables. The concave-convex property is non-trivial and important, because the bit thread derivations will be via Lagrange dualisation.

We start by listing three surface-based holographic prescriptions for calculating the boundary entropy $S(A)$, two of which are novel. We are going to show that these prescriptions are equivalent to each other and the QES formula~\eqref{eq:QESform}. 
\begin{enumerate}

\item Quantum maximin:
\bne S_- =  \adjustlimits \sup_{\sigma \in \sSA} \inf_{r \in \sR_A} \!\!S_{\textrm{gen}}(\sigma \cap r) \,.
\ene
\end{enumerate}
This is known to be equivalent to the QES prescription, i.e. $S_- = S(A)$~\cite{Akers:2019lzs}. The proof assumes the QFC.
\begin{enumerate}[resume]
\item Quantum minimax:
\bne S_+ = \adjustlimits\inf_{r\in \sR_A} \sup_{\sigma\in \sSA} \,S_{\textrm{gen}}(\sigma \cap r)\,. \label{eq:qminimax}
\ene
\end{enumerate}
Minimax and maximin are related by a simple swap in the order of optimisation over slices and homology regions. This swap may seem like a small change, but it is non-trivial that quantum maximin and minimax are equivalent. The proof will be given in Sec.~\ref{sec:surpr}.

Neither the quantum maximin nor minimax prescriptions are concave-convex optimisation problems (in $\sigma$ and $r$), because $\sR_A$ and $\sSA$ are not convex sets,
and this is an obstacle to Lagrange dualising them to equivalent bit thread prescriptions. 

This naturally leads us to our next surface-based prescription. It is an optimisation problem over measures on the sets of slices and homology regions, and it \textit{is} concave-convex in those measures.
\begin{enumerate}[resume]
\item Relaxed quantum maximin:\footnote{We implicitly specify the sets on which the probability measures are functions through the specification of the integration domains (for example, as in~\eqref{eq:fdef}).} 
\bne S_c = \adjustlimits \sup_{\kappa} \inf_{\rho} f(\rho, \kappa)\label{eq:coqmax} \ene
with
\bne \label{eq:fdef}f(\rho, \kappa) :=  \frac{1}{4G_N}\int_{M} \!\!\sqrt{g}\, |d\phi_{\kappa} \wedgedot d\psi_{\rho} | + \int_{\sSA} d\kappa (\sigma) \int_{\sR_A}\!\!\!  d\rho (r)  S_b (\sigma \cap r)  \,,
\ene
where the scalar function $\psi_\rho$ and $\phi_\kappa$ are as defined in~\eqref{eq:phipsidefs}.
\end{enumerate}
 
The wedgedot function $|\, \wedgedot \,|$ in~\eqref{eq:fdef} is a scalar concave-convex function on a pair of 1-forms in $j^+ \times T^*$, with $j^+ \subset T^*$ the set of future-directed causal 1-forms. The definition of the wedgedot function is, for $(W,X) \in j^+ \times T^*$,
\bne |W \wedgedot X| := \max \{ |W\cdot X|,|W\wedge X|\} = \begin{cases} |W \cdot X|, &X^2 \leq 0\\
|W\wedge X|, &X^2 \geq 0 \end{cases} \,.
\ene
The reason we need the wedgedot is that the 2-form norm $|d\phi \wedge d\psi|$, which appears in the coarea formula
\bne \int dt \int ds \,\text{area}(\sigma_t\cap \tau_s) = \int_{M}\sqrt{g} |d\phi \wedge d\psi|\,, 
\label{eq:coarea} \ene
where $\sigma_t = \phi^{-1}(t)$ and $\tau_s = \psi^{-1}(s)$ are the level sets of $\phi$ and $\psi$, is not concave in $d\phi$ when $d\psi$ is timelike, which is a problem for Lagrange dualisation. The wedgedot is the unique concave-convex function on $j^+ \times T^*$ that equals $|W\wedge X|$ for $X^2 \geq 0$~\cite{Headrick:2022nbe}. The wedgedot has other properties~\cite{Headrick:2022nbe}, which we will state when needed. 

The following properties of the relaxed quantum maximin prescription~\eqref{eq:coqmax} make it concave-convex:
\begin{enumerate}
	\item The area term $|d\phi_{\kappa} \wedgedot d\psi_{\rho}|$ is concave in $\kappa$ and convex in $\rho$. 
	\item The bulk entropy term in $f$ is also concave-convex, since it is linear in the measures. 
	\item The variables we are optimising over are convex sets, since the set of all possible probability measures is a convex set.
\end{enumerate}

Note that, unlike the unrelaxed formulas, and because the objective function is concave-convex, it follows that the order of minimisation and maximisation in the relaxed quantum maximin prescription, Eq.~\eqref{eq:coqmax}, does not affect its optimal value, i.e.\ maximin equals minimax.%
\footnote{Sion's minimax theorem gives a set of sufficient conditions for a maximin and minimax pair to be equivalent, which includes a requirement that the objective function is concave-convex~\cite{sion1958general}. The theorem also requires a continuity assumption on the objective function and a compactness assumption on the domain of at least one of the variables. Since we are not aiming to be rigorous, we will not delve into these analytic issues.
}

In subsection~\ref{sec:surpr}, we will prove that all the above surface-based prescriptions are equivalent, i.e.\ that $S_+ = S_c = S_- (= S(A))$. 

\subsubsection{Other surface-based prescriptions} \label{sec:OTsupr}
Before doing so, let us briefly discuss two more surface-based prescriptions. We will not dualise these two to bit thread prescriptions, but they are of sufficient interest to discuss here.

The first we call the level-set relaxed
 quantum maximin prescription:
\bne \adjustlimits \sup_\phi \inf_\psi  \left( \frac{1}{4G_N} \int_{M} \sqrt{g} |d\phi \wedgedot d\psi| + \int_{0}^{1}\!\! dt \int_{0}^{1} \!\!ds S_b (\sigma_t \cap r_s) \right).
\label{eq:ppqmm} \ene
One can rewrite this in a way that makes its relation to a smearing out of $\Sgen$ more apparent:
\bne \adjustlimits \sup_\phi \inf_\psi  \int_{0}^{1}\!\! dt \int_{0}^{1} \!\!ds \left(\frac{\textrm{area}'(\sigma_t, \tau_s)}{4G_N} + S_b (\sigma_t \cap r_s)\right),
\label{eq:ppqmm2} \ene
where area$'$ is the modification of the area functional that arises if one replaces the ordinary wedge product in the coarea formula~\eqref{eq:coarea} with the wedgedot. Its definition is
\bne \text{area}'(\sigma,\tau) := \int_{\sigma \cap \tau} \sqrt{h} \times \begin{cases} 1 & \text{$\tau$ timelike or null at $\sigma$}\\
\coth \chi & \text{$\tau$ spacelike at $\sigma$}
\end{cases} \,,
\ene
where $\chi$ is the dihedral boost angle between $\sigma$ and $\tau$.

The classical limit of the prescription~\eqref{eq:ppqmm} is concave-convex (concave in $\phi$ and convex in $\psi$), and equivalent to the HRT prescription~\cite{Headrick:2022nbe}. However, the addition of the bulk entropy term in~\eqref{eq:ppqmm} spoils the concavity with respect to $\phi$, as we show in App.~\ref{sec:noc}. As with unrelaxed quantum maximin/minimax, this is an issue, because it means that the maximin and minimax versions of this prescription are not guaranteed to be equivalent,
and that the U-flow prescription we could derive by dualising on $\phi$ would not be guaranteed to have zero duality gap. 

The next prescription \textit{is} concave-convex (because it is linear in the measures):
\bne \adjustlimits \inf_{\rho} \sup_{\kappa}\int_{\sR_A}\!\!\!  d\rho (r) \int_{\sSA} d\kappa (\sigma) S_{\textrm{gen}}(\sigma \cap r) \,.
\label{eq:mmasg} \ene 
This means that its maximin and minimax prescriptions are equivalent, and that the dual optimisation problems will have zero optimal duality gap (i.e. be equivalent). 
Note that the objective function in~\eqref{eq:mmasg} \emph{is} an averaged generalised entropy, in contrast to~\eqref{eq:coqmax}.%
\footnote{
The function $f$ in~\eqref{eq:coqmax} is not an averaged generalised entropy because, for general measures,  
\bne \int_{\sR_A}\!\!\!  d\rho \int_{\sSA} d\kappa |d\chi_\sigma \wedgedot  d\chi_r| \neq \left|  \left( \int_{\sSA} d\kappa \, d\chi_\sigma \right) \wedgedot \left( \int_{\sR_A}\!\!\!  d\rho  \, d\chi_r \right)\right|. \ene
}
We leave the exploration of the prescriptions dual to~\eqref{eq:mmasg} to future work, but we expect them to be the covariant generalisations of the cutoff-independent flow prescriptions of~\cite{Headrick:2025awv}.

\subsubsection{Proofs of mutual equivalence} \label{sec:surpr}
Now we will prove that the three surface-based prescriptions for holographic entanglement entropy we introduced in Sec.~\ref{QuantumMaximin} are equivalent to each other.
This is important because it guarantees that the bit thread prescriptions that we will derive from dualising the relaxed quantum maximin prescription are QES-equivalent.
To prove equivalence, i.e. $S_+ = S_c = S_- (= S(A))$, it suffices to prove the following three relations: $S_+ = S_-$, $S_+ \geq S_c$ and $S_c \geq S_-$.

\addtocontents{toc}{\protect\setcounter{tocdepth}{2}}
\subsubsection*{Quantum maximin equals quantum minimax ($\mathbf{S_+ = S_-}$) }
\addtocontents{toc}{\protect\setcounter{tocdepth}{3}}

The classical maximin and minimax prescriptions are equivalent when the spacetime obeys AdS boundary conditions and classical focussing ($\theta' \leq 0$)~\cite{Headrick:2022nbe}. The role of classical focussing is to ensure that, amongst candidate surfaces contained within the entanglement wedge boundary timesheet, the HRT surface is area-maximising. 
The argument for the equivalence of the classical prescriptions straightforwardly extends to \textit{quantum} maximin and minimax (so we will omit details). 
All that needs to be changed is the replacement of area with generalised entropy, and classical focussing ($\theta' \leq 0$) with the QFC ($\Theta' \leq 0$).
The QFC plays the same role that the null energy condition (NEC) does in the classical case: the NEC ensures that the HRT surface is area-maximising on the entanglement wedge horizon $\tau (\cW(A))$, while the QFC ensures that the QES is $\Sgen$-maximising.

\addtocontents{toc}{\protect\setcounter{tocdepth}{2}}
\subsubsection*{\texorpdfstring{$\mathbf{S_c \geq S_-}$}{Convex greater than minimax}} \label{sec:ScGtrSm}
\addtocontents{toc}{\protect\setcounter{tocdepth}{3}}

To prove that $S_c \geq S_-$, we first note that
\bne S_c = \adjustlimits \sup_{\kappa} \inf_{\rho} f(\rho, \kappa) \geq  \adjustlimits \sup_{\sigma\in \sSA} \inf_{\rho} f(\rho, \kappa)\big |_{\kappa = \delta_\sigma} \ene
(where $\delta_x$ is shorthand for a $\delta-$function with support at $x$ in the relevant domain), since $\delta$-function measures are a subset of all probability measures, so restricting to them cannot increase the supremum. This simplification to maximising over slices will allow us to use properties of bulk entropies of slice subregions.
We have that
\bne \inf_{\rho} f(\rho, \kappa)\big |_{\kappa = \delta_\sigma} = \inf_{r\in \sR_A} f(\rho, \kappa)\big |_{\rho = \delta_r, \,\kappa = \delta_\sigma} \,. \ene
This follows from the SSA of bulk entropies (see the ``disentangling" lemmas in~\cite{Headrick:2025awv} for proofs). Lastly, we use that 
\bne \adjustlimits \sup_{\sigma \in \sSA}  \inf_{r\in \sR_A}  \!\!f(\rho, \kappa)\big |_{\rho = \delta_r, \,\kappa = \delta_\sigma} = S_- \, . \ene
Together, these imply that $S_c \geq S_-$.

\addtocontents{toc}{\protect\setcounter{tocdepth}{2}}
\subsubsection*{\texorpdfstring{$\mathbf{S_+ \geq S_c}$}{Maximin greater than convex}} \label{eq:SpgteSc}
\addtocontents{toc}{\protect\setcounter{tocdepth}{3}}

Let $\tau_A$ be the timesheet of the entanglement wedge $\cW(A)$. Since $\tau_A$ is an optimal timesheet for the quantum minimax prescription, we have, by definition, that
\bne S_+ = \sup_{\sigma\in \sSA} S_{\textrm{gen}} (\sigma \cap \cW(A))\,. \label{eq:Spsim} \ene
We also have that
\bne S_c \leq \sup_{\kappa} f(\rho ,\kappa)\big|_{\rho = \delta_{\cW(A)}} \label{eq:Scbou} \ene
because the measure $\rho = \delta(r-\cW(A))$ is not necessarily optimal for $S_c$.

Next, a useful lemma is that
\bne \int_{\sSA} d\kappa (\sigma)\, S_{\textrm{gen}} (\sigma \cap \cW(A)) = f(\rho,\kappa) \big|_{\rho = \delta_{\cW(A)}}\,.
\label{eq:lemm1} \ene

To prove this, we note that $\tau_A$ is piecewise null and use that $|W \wedgedot X|$ is linear in $W \in j^+$ if $X$ is null~\cite{Headrick:2022nbe}, which implies that the area piece of $f$ satisfies
\bne \left|\int_{\sSA} d\kappa(\sigma) \, n_\sigma \wedgedot n_{\tau_A}\right| = \int_{\sSA} d\kappa (\sigma) |n_\sigma \wedgedot n_{\tau_A}| \,.
\ene
This is sufficient to prove~$S_c \leq S_+$, because~\eqref{eq:Scbou} and~\eqref{eq:lemm1} together give us%
\footnote{The equality follows from~\eqref{eq:Spsim} and the fact that maximising the weighted average of a function over probability measures on a set is equivalent to maximising the function over the set, i.e.\ $\sup_\mu \int_X  f(x) d\mu(x) = \sup_{x\in X} f(x)$.}
\bne S_c \leq \sup_{\kappa} \int_{\sSA} d\kappa (\sigma) S_{\textrm{gen}} (\sigma \cap \cW(A)) = S_+\,.\ene

Note that there is an asymmetry between the proofs of $S_+ \geq S_c$ and $S_c \geq S_-$, in the sense that we needed SSA of bulk entropies to prove the latter but not the former. The reason for the asymmetry --- why the above proof of $S_+ \geq S_c$ does not work for proving that $S_c \geq S_-$ ---
is that $|W\cdot X|$ is linear in $W \in j^+$ when $X^2 = 0$, but $|W \cdot X|$ is not linear in $X \in T^*$ when $W^2 = 0$.

\subsection{Lightning review of Lagrange dualisation}
We will use Lagrange dualisation to derive the quantum covariant bit thread prescriptions of Sec.~\ref{sec:QCBIT}. 
Lagrange dualisation consists of three steps:
\begin{enumerate}
    \item Take a constrained minimisation problem, referred to as the primal problem.
    \item Construct a Lagrangian from the primal problem by adding Lagrange multiplier terms to enforce the constraints.
    \item Minimise the Lagrangian over the primal variables to get the dual function. The dual problem is the dual function maximised over the Lagrange multiplier variables.
\end{enumerate}
If the primal and dual problems are equivalent, then strong duality is said to hold, otherwise there is a non-zero duality gap. A sufficient condition for strong duality to hold in convex optimisation problems is Slater's condition: there exists a strictly feasible point in the domain of the primal problem. 

Constraints in the dual problem arise from restricting the dual function to its feasible domain: where it is greater than minus infinity.
For readers interested in further details on Lagrange duality and convex optimisation, see~\cite{Boyd:2004fnq}.

\subsection{Loose quantum V-flows} \label{sec:loosV}

The primal problem that we will dualise is the relaxed quantum maximin prescription~\eqref{eq:coqmax}, except with $d\psi_{\rho}$ replaced by a 1-form $w$ and the equality constraint $w = d\psi_{\rho}$. We make this replacement because different primal problems, even if trivially equivalent, can give different dual problems. The flow field $V$ will appear as the Lagrange multiplier of the $w = d\psi_{\rho}$ constraint. 
Our primal problem is
\bne \begin{split}
&\inf_{\rho} \left( \frac{1}{4G_N} \int_{M} \!\! \sqrt{g} |d\phi_{\kappa} \wedgedot w | + \int_{\sSA} d\kappa (\sigma) \int_{\sR_A}\!\!\!  d\rho (r)  S_b (\sigma \cap r)\right )\\
&\qquad \qquad \text{subject to} \quad w = \int_{\sR_A}\!\!\!  d\rho \, d \chi_{r} \text{ and } \int_{\sR_A}\!\!\!  d\rho = 1\,. \end{split}\label{eq:wintr}
\ene

We have also not included in our primal problem the $\kappa$-maximisation aspect of relaxed quantum maximin~\eqref{eq:coqmax}. Since we are only dualising on $\rho$, not $\kappa$, it would only come along for the ride. We can and will reintroduce the $\kappa$-maximisation at the end, in the dual problem, to get our V-flow prescription equivalent to quantum maximin.

The Lagrangian for the primal problem~\eqref{eq:wintr} is
\bne \begin{split} L (\rho, w,V,\lambda) = \int_{M} \sqrt{g} \left(  \frac{1}{4G_N}|d\phi_{\kappa} \wedgedot w|  +  V \cdot \left( \int_{\sR_A}\!\!\!  d\rho \, d \chi_{r} -w\right) \right) + \lambda \left( 1- \int_{\sR_A}\!\!\!  d\rho\right) \\
+ \int_{\sSA} d\kappa \int_{\sR_A}\!\!\!  d\rho \, S_{b} (\sigma \cap r)\,. \label{eq:Vlang} \end{split}\ene
To derive the problem dual to relaxed quantum maximin, the recipe states to minimise the Lagrangian~\eqref{eq:Vlang} over the primal variables ($w$ and $\rho$).
The infimum over $w$ is only finite if the following constraint on $\phi_{\kappa}$ and $V$ holds~\cite{Headrick:2022nbe}
\begin{gather} 
\inf_w \int_{M} \sqrt{g} \left(  \frac{1}{4G_N}|d \phi_{\kappa}\wedgedot w| - V\cdot w\right ) > -\infty \\[-0.5em]
\Updownarrow \nonumber \\[-0.5em]
\frac{d\phi_{\kappa}}{4G_N}  \pm V \in j^+ 
\label{eq:firstC}
\end{gather}

This is the same norm bound constraint as in the classical prescription. The reason for this is that the $w$-dependent terms in the Lagrangian~\eqref{eq:Vlang} do not depend on the bulk entropies. 

The minimisation over the measure $\rho$ gives a constraint on the flow field $V$ and the scalar $\lambda$ in terms of bulk entropies. The constraint holds for all homology regions $r$:%
\footnote{We use the convention that normal covectors on the boundaries of spacetime regions are inward-pointing. This is why our statement of Stokes' theorem has a relative sign difference in comparison to the standard convention. This also means that $d\chi_r$ and the normal covector $n_{\tau(r)}$ have the opposite orientation.}
\begin{gather}
\inf_{\rho} \left( \int_{M} \sqrt{g} \left( V \cdot \int_{\sR_A}\!\!\!  d\rho \, d \chi_r \right) - \lambda \int_{\sR_A}\!\!\!  d\rho  + \int_{\sSA} d\kappa \int_{\sR_A}\!\!\!  d\rho \, S_{b} (\sigma \cap r) \right ) >-\infty \\
\Updownarrow \nonumber\\[-0.5em]
\forall r\in \sR_A: -\int_{\tau(r)} *V - \lambda + \int_{\sSA} d\kappa (\sigma) S_b (\sigma \cap r) \geq 0\,. \label{eq:lambin}
\end{gather}
We get an inequality in the constraint, rather than an equality, because we have non-negativity of the measure $\rho$ as an implicit constraint.

So, the resulting dual problem is
\bne \begin{split} \sup_V \sup_\lambda \, (\lambda)\label{eq:dual1} \end{split}\ene
subject to~\eqref{eq:firstC} and~\eqref{eq:lambin}.

We can eliminate $\lambda$ from this problem easily, using the inequality constraint~\eqref{eq:lambin} to get
\bne \begin{split} \label{eq:Vatnew} & \adjustlimits \sup_V \inf_{r\in \sR_A} \left(-\int_{\tau(r)} \!\! *V + \int_{\sSA} d\kappa (\sigma) S_b (\sigma \cap r) \right)  \end{split} \ene
subject to the norm bound~\eqref{eq:firstC}.

We will eliminate $r$ from this problem to simplify further. To do so, we first need to argue that there always exists an optimal flow for the problem~\eqref{eq:Vatnew} that obeys certain properties. These properties can then be imposed as additional constraints in the problem, as they will not affect the optimal value, and these constraints will allow us to simplify~\eqref{eq:Vatnew}.

First, we will need Stokes' theorem in the form 
\bne -\int_{\tau(r)}\!\!\!\!\!  *V = \int_{r} \, d *V + \int_{(\cI^+ \cup \cI^-)\cap r} \!\!\!\!\! *V + \int_{D(A)} \!\!\!\!\!\!\!\! *V\,. \label{eq:Stke1} \ene
Using Stokes' theorem, we rewrite~\eqref{eq:Vatnew} as
\bne \label{eq:Vat2}\adjustlimits \sup_V \inf_{r\in \sR_A} \left(\int_{r} d *V + \int_{(\cI^+ \cup \cI^-)\cap r}\!\!\!\!\!\! *V + \int_{D(A)}\!\!\!\!\!\!\!\! *V + \int_{\sSA} d\kappa (\sigma) S_b (\sigma \cap r) \right)  
\ene
subject to the norm bound~\eqref{eq:firstC}.

Next, we will argue that optimal V-flows for the problem~\eqref{eq:Vat2} always satisfy
\bne *V|_{\cI^+ \cup \cI^-} = 0 \,.
\label{eq:vIcon} \ene
For fixed $V$,
\begin{enumerate} 
\item If $\int_{N(p)} *V >0$ for an infinitesimal neighbourhood $N(p)$ of a point $p \in \cI^+ \cup \cI^-$, then we can decrease $\int_{(\cI^+ \cup \cI^-) \cap r}*V$ by excluding $p$ from $r$.
\item If $\int_{N(p)} *V < 0$, then we can decrease $\int_{(\cI^+ \cup \cI^-) \cap r}*V$ by including $p$ in $r$.
\end{enumerate}
So, from the $\sup_V$'s perspective, having $V \neq 0$ on $\cI^+ \cup \cI^-$ can only be disadvantageous. Therefore, we are free to impose~\eqref{eq:vIcon} as a constraint on all $V$.

Using this result, we can rewrite~\eqref{eq:Vat2} as
\bne \begin{split} \label{eq:Vat3}\sup_V 
\left( \int_{D(A)}\!\!\!\!\!\!\!\! *V + \mathcal{F}(V) \right)&  
\quad \text{ over } V \\
\text{ subject to: }  &\frac{d\phi_{\kappa}}{4G_N} \pm V \in j^+, \\
& *V|_{\cI^+ \cup \cI^-} = 0  \, ,
\end{split} \ene
where
\bne \label{eq:mco2}\cF(V):= \inf_{r \in \sR_A} \left( \int_{r}  d*V + \int_{\sSA} d\kappa (\sigma) S_b (\sigma \cap r)\right)  .\ene

Next, we will argue that there is guaranteed to exist an optimal flow which satisfies $\cF(V^*) = 0$. This will allow us to remove $\cF(V)$ from the objective function in~\eqref{eq:Vat3}, in exchange for an additional constraint.
The essence of the argument is that, from an optimal flow $V^*$ (optimal for~\eqref{eq:Vat3}) that does not satisfy $\cF(V^*) = 0$, we can construct a new optimal flow $V_{\textrm{new}}^*$ which does.
This argument is similar to one used in the derivation of the static quantum flow prescriptions~\cite{Headrick:2025awv}. 
\begin{enumerate}
    \item First, suppose that we have a $V^*$ for which $\cF(V^*) <0$. 
    This inequality implies that there must be more sinks than sources for $V^*$ in $r^*$, because $S_b \geq 0$. Then, from $V^*$, we can construct a $V_{\textrm{new}}^*$ for which $\cF(V_{\textrm{new}}^*)=0$.
    The new flow $V_{\textrm{new}}^*$ is constructed from the original flow by the removal of sinks in $r^*$ and their attached field lines. This removal does not lead to a violation of the norm bound constraint, because $V_{\rm{new}}^* = 0$ along those old field lines, and since $\frac{d\phi_{\kappa}}{4G_N} \pm V^* \in j^+$ is satisfied, then so is $\frac{d\phi_{\kappa}}{4G_N} \pm 0 \in j^+$.
    Those field lines have to end either on the boundary region $D(A)$ or in the bulk region $r^*$, because otherwise the flux $-\int_{\tau(r^*)} *V$ in~\eqref{eq:Vatnew} would not have been optimal. So, the removal does not affect $-\int_\tau *V$, and the new flow is therefore still optimal. The timesheet $\tau(r^*)$, which was optimal for $V^*$, is still optimal for $V_{\textrm{new}}^*$, because the flux through $\tau (r^*)$ is the same for $V^*$ and $V_{\textrm{new}}^*$, and the flux through other timesheets can only increase as sinks are removed.
    \item Next, we consider the opposite case, a $V^*$ for which $\cF(V^*) >0$. 
    To construct a $V^*_{\textrm{new}}$ satisfying $\cF(V^*_{\textrm{new}})=0$ from this $V^*$, we simply add a sufficient number of flow lines from $D(A)$ to sinks that are infinitesimally into the bulk.
    It is a reasonable assumption that we can modify $V^*$ this way
    without violating the norm bound, because the bulk volume diverges near the AdS boundaries. The homology region $r^*$ remains optimal, because we have not changed the flux through $\tau(r^*)$.
\end{enumerate}

We have argued that there always exists an optimum $V^*$ for which $\cF(V^*)=0$, and this allows us to impose $\cF(V)=0$ as an additional constraint in~\eqref{eq:Vat3} (without affecting the optimal value), which allows us to remove $\cF(V)$ from the objective function in~\eqref{eq:Vat3}. Doing so and restoring the $\kappa$-maximisation gives us the  end result, our loose quantum V-flow prescription:\fn{We have used that $\min_{x,y} f(x)$ subject to $g(x,y) = 0$ is equivalent to $\min_x f(x)$ subject to $\exists y_0: g(x,y_0) =0$.}
\begin{equation}
	\label{eq:qcVexists}
		\begin{aligned}
			S(A) =  \sup_{\kappa , V}  \int_{D(A)}\!\!\!\!\! *V& \\[4pt]
			\qquad \qquad \text{subject to:} \quad 
			& *V\big|_{\cI^+ \cup \cI^-} = 0 \\[4pt]
			&\frac{d\phi_{\kappa}}{4G_N} \pm V \in j^+\\[2pt]
			 &\forall r\in \sR_A:\; 
			\int_{r} d{*V}
			+ \int_{\sSA} d\kappa(\sigma)\, S_b(\sigma \cap r)
			\ge 0\,.
		\end{aligned}
\end{equation}

We have replaced $\cF(V) = 0$ with an equivalent inequality constraint which does not require prior knowledge of an optimal homology region. 
Only $V$ appears in the objective function, not the measure $\kappa$.

There is an equivalent form of~\eqref{eq:qcVexists} that is closer to the classical covariant V-flow prescription.
We want to replace the maximisation
over the measure $\kappa$ with a maximisation over a scalar field $\phi$. The obstruction is that generic measures $\kappa$ have support on slices which are not nested. But, since there exists an optimum $\kappa$ which only has support on a single slice, we can impose this nesting property on $\kappa$ without affecting the optimum, and rewrite~\eqref{eq:qcVexists} as
\begin{equation}
\label{eq:qcV2}
\begin{aligned}
\sup_{\phi, V} & \int_{D(A)} \!\!\!\!\! *V \\[4pt]
\text{subject to} \quad 
& *V\big|_{\mathcal I^- \cup \cI^+} = 0, \\[4pt]
& \phi|_{\mathcal{I}^-} \geq 0\qquad \phi|_{\mathcal{I}^+} \leq 1\\
&\frac{d\phi}{4G_N} \pm V \in j^+ \\[2pt]
& \forall r \in \sR_A :\;
\int_{r}  d{*V}
+ \int_{0}^{1} \!\! dt\, S_b(\sigma_t \cap r)
\ge 0\,.
\end{aligned}
\end{equation}
where $\sigma_t$ are the level sets of $\phi$. 
A disadvantage of writing the V-flow prescription this way is that it is not concave in $\phi$ (in contrast to~\eqref{eq:qcVexists}, which is concave in $\kappa$).

\subsection{Strict quantum V-flows}

Here we will show that the strict quantum V-flow prescription is QES-equivalent. We will dualise from the flow prescription to the surface-based prescription; this is the opposite direction to that we took for the loose prescription in Sec.~\ref{sec:loosV}.

We make only the strict quantum V-flow divergence constraint explicit in the primal problem: 
\bne
\begin{split}
\sup_{V,\,\kappa} \quad 
& \int_{D(A)}\!\!\! *V \\
\text{subject to:} 
& \quad \forall\, r \in \sR:\;
\left|\int_{r} d{*V}\right| \le \int_{\sSA} d\kappa(\sigma) S_b(\sigma \cap r)\,.
\label{eq:svflo}
\end{split}
\ene
For the purposes of dualisation, the norm bound $\frac{d\phi_{\kappa}}{4G_N} \pm V \in j^{+}$ and boundary condition $*V\big|_{\mathcal{I}^+\cup\mathcal{I}^-} = 0$ are left as implicit constraints. We will dualise on $V$, leaving the supremum over $\kappa$ untouched.

The Lagrangian for the primal problem~\eqref{eq:svflo} is, after integrating the $d*V$ divergence term by parts,
\bne \begin{split} L(V,\kappa, \rho^{(\pm)}) = \int_{D(A)} \!\!(1+ \psi_- - \psi_+)\, {*V} + &\int_{D(A^c)}\!\!(\psi_- - \psi_+)\, {*V} + \int_{M} \!\!\sqrt{g}\, d(\psi_- - \psi_+) \cdot V  \\
 &+\int_{\sSA}d\kappa(\sigma)\int_{\sR} (d\rho^{(+)} (r)  + d\rho^{(-)}(r))  S_b(\sigma \cap r)\,,
\label{eq:sctVL} \end{split} \ene
where 
\bne \psi_\pm := \int_{\sR} d\rho^{(\pm)} (r) \chi_{r} \, . \label{eq:pside} \ene
This Lagrangian has a pair of Lagrange multiplier measures 
$\rho^{(\pm)}$, because the strict divergence constraint in~\eqref{eq:svflo} is of the form $|X| \leq Y$, which is equivalent to the pair of constraints $X \leq Y$ and $-X \leq Y$.

While $\rho^{(\pm)}$ are measures (they have a definite sign because they are enforcing inequality constraints), they are not at this stage~\textit{probability} measures, i.e.\ normalised to one. The normalisation will come as a result of the dualisation.

Next, we maximise the term in the Lagrangian~\eqref{eq:sctVL} that is an integral over the spacetime $M$. To do so, we use the identity~\cite{Headrick:2022nbe}
\bne \sup_{\left \{(\phi,V)|\frac{d\phi}{4G_N} \pm V \in j^+ \right \}} ( d\psi \cdot V) \equiv \frac{1}{4G_N} \sup_{\{\phi| d\phi \in j^+\}}|d\phi \wedgedot d\psi|\,. \ene

This gives us the following dual problem:

\bne \begin{split} \adjustlimits \sup_{\kappa}\inf_{\rho^{(\pm)}} &\left( \frac{1}{4G_N}\int_M \!\! \sqrt{g} |d\phi_{\kappa} \wedgedot d(\psi_- - \psi_+)| \right.\\
&\qquad\qquad\qquad+\left. \int_{\sSA}d\kappa(\sigma) \int_{\sR} (d\rho^{(+)} (r) + d\rho^{(-)} (r)) S_b(\sigma \cap r)\right) \\
\qquad \text{subject to: } 
&(\psi_- - \psi_+)|_{D(A)} = -1 \\
&(\psi_- - \psi_+)|_{D(A^c)} = 0\,. 
\label{eq:strictVflow} \end{split}\ene
We will show that~\eqref{eq:strictVflow} is equivalent to the QES prescription (in its convex-relaxed quantum maximin form~\eqref{eq:coqmax}). We will do so by showing that there is guaranteed to exist optimal measures $\rho^{(\pm)}$ for which
\begin{enumerate}
    \item $\rho^{(-)} =0$
    \item $\rho^{(+)}$ has support on only $\mathscr{R}_A \subset \mathscr{R}$
    \item $\int_{\cR_A}\! d\rho^{(+)}(r) =1$, i.e.\ $\rho^{(+)}$ is a probability measure.
\end{enumerate}
For that, we need an uplift of the disentangling lemmas of~\cite{Headrick:2025awv} to our covariant setup. The disentangling lemmas allow us to restrict to measures $\rho^{(\pm)}$ with non-overlapping support on nested homology regions.
Since many of the ingredients in the proof are the same as in the static setup, we will just sketch how the key part of the proof uplifts. 

\begin{figure}[htbp]
    \centering
    \includegraphics[width=0.8\linewidth]{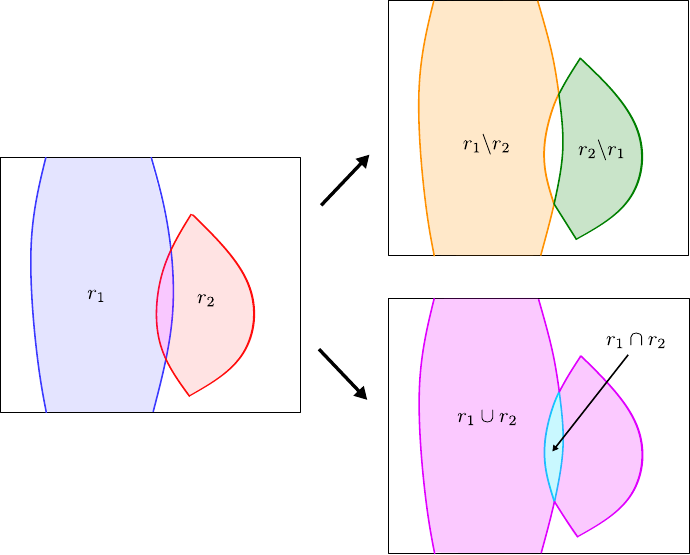}
    \caption{A typical pair of homology regions $r_1$ and $r_2$, and how we map them to either disjoint or nested regions. 
    }
    \label{fig:CovDistengtangling}
\end{figure}
\begin{enumerate}
	\item \textbf{Overlapping regions.} Consider a typical pair of homology regions $r_1$ and $r_2$, as depicted in Fig.~\ref{fig:CovDistengtangling}, which are neither disjoint nor nested, and let $\rho^{(+)} (r_1) \neq 0$ and $\rho^{(-)} (r_2) \neq 0$. Then, from the old measures $\rho^{(\pm)}$, we map to new measures
	\bne \rho^{(+)}(r_1) \mapsto \rho^{(+)}(r_1) -\alpha \label{eq:musta}\ene
	\bne \rho^{(-)}(r_2) \mapsto \rho^{(-)}(r_2) -\alpha \ene
	\bne \rho^{(+)}(r_1 \backslash r_2) \mapsto \rho^{(+)} (r_1 \backslash r_2) + \alpha \ene
	\bne \rho^{(-)}(r_2 \backslash r_1) \mapsto \rho^{(-)} (r_2 \backslash r_1) + \alpha\ene
	where $\alpha := \min \{ \rho^{(+)} (r_1), \rho^{(-)} (r_2) \}$. Roughly speaking, we have shifted the support of the measures from the overlapping regions $r_1$ and $r_2$ to the disjoint regions $r_2 \backslash r_1$ and $r_1 \backslash r_2$. 
	We repeat this procedure until $\rho^{(+)}$ and $\rho^{(-)}$ have disjoint support (for a careful argument in the static case, which we expect to generalise, see~\cite{Headrick:2025awv}).
	Under this map, the objective function in~\eqref{eq:strictVflow} is non-increasing, because $(\psi_- - \psi_+)$ is invariant, and because of the following weak monotonicity property of von Neumann entropies~\cite{lieb1973proof}
	\bne S(X) + S(Y) \geq S(X\backslash Y) + S(Y\backslash X) \,.\ene
	Therefore, we can restrict to measures $\rho^{(\pm)}$ with disjoint support.

    \item \textbf{Nested regions.} By a similar argument, we can restrict the measures $\rho^{(\pm)}$ to have support on nested sets of regions. Suppose the contrary, that one of $\rho^{(\pm)}$ (let us take it to be $\rho^{(+)}$), has support on non-nested regions, like $r_1$ and $r_2$ in Fig.~\ref{fig:CovDistengtangling}. Then we change the measures $\rho^{(\pm)}$ by transferring their support from $r_1$ and $r_2$ to the nested pair of regions $r_1 \cup r_2$ and $r_1 \cap r_2$:
	\bne \rho^{(+)}(r_1) \mapsto \rho^{(+)}(r_1) -\alpha \ene
	\bne \rho^{(+)}(r_2) \mapsto \rho^{(+)}(r_2) -\alpha \ene
	\bne \rho^{(+)}(r_1 \cup r_2) \mapsto \rho^{(+)} (r_1 \cup r_2) + \alpha \ene
	\bne \rho^{(+)}(r_2 \cap r_1) \mapsto \rho^{(+)} (r_2 \cap r_1) + \alpha \label{eq:muend}\ene    
	where $\alpha := \min \{ \rho^{(+)} (r_1), \rho^{(+)} (r_2) \}$. 
	Under this map, the scalar function $\psi_+$ is unchanged, and the entropy is non-increasing, by SSA. We repeat this procedure until all the regions $r$ for which $\rho^{(+)}$ has support are nested.
\end{enumerate}

We have shown that we can restrict to measures $\rho^{(\pm)}$ which do not overlap and have support only on nested regions. Then, since non-zero $\psi_-$ can only contribute positively to the objective function in~\eqref{eq:strictVflow}, and not help with satisfying the $(\psi_- - \psi_+)|_{D(A)} = -1$ constraint, it follows that the optimum $\rho^{(-)}$ is $0$. It also follows that optimal measures $\rho^{(+)}$ are probability measures on $\mathscr{R}_A$, because of the $\psi_+|_{D(A)} = 1$ constraint; part of the argument is that this constraint requires that $\int_{\mathscr{R}} \rho^{(+)} \geq 1$.

\subsection{Quantum U-flows}
The primal problem that we will dualise is the relaxed quantum minimax%
\footnote{Recall that relaxed quantum maximin equals relaxed quantum minimax.} prescription~\eqref{eq:coqmax}, with $d\phi_\kappa$ in the area term $|d\phi_\kappa \wedgedot d\psi_{\rho}|$ replaced by $d\phi$ and the equality constraint $d\phi = d\phi_\kappa$. $U$ will arise as the Lagrange multiplier for this equality constraint.

Here is the primal problem:
\bne \begin{split} &\sup_{\phi, \kappa} \left(\frac{1}{4G_N} \int_{M}\!\! \sqrt{g} |d\phi \wedgedot d\psi_{\rho}| + \int_{\sSA} d\kappa (\sigma) \int_{\sR_A}\!\!\!  d\rho (r) S_b (\sigma \cap r) \right), \\
	\textrm{subject to: }  &d\phi =  \int_{\sSA} d\kappa (\sigma)\, d\chi_{J^+ (\sigma)}\\
    &\int_{\sSA} d\kappa =  1\,. \label{eq:Uprim}\end{split} \ene
The non-negativity $\kappa \geq 0$ of the measure on slices, and future-directedness of $d\phi$
are left as implicit constraints. 

We have not included in our primal problem the $\rho$-minimisation aspect of relaxed quantum minimax~\eqref{eq:coqmax}. Since we are only dualising on $\kappa$, not $\rho$, it would only come along for the ride. The measure $\rho$ is taken to be fixed in~\eqref{eq:Uprim}.  We will reintroduce $\rho$-minimisation at the end to get our U-flow prescription.

The Lagrangian for this primal problem is 
\bne \begin{split} L(\phi,\kappa,\lambda,U) = \frac{1}{4G_N} \int_{M}\!\! \sqrt{g} |d\phi \wedgedot d\psi_{\rho}|  + \int_{\sSA} d\kappa(\sigma)  \int_{\sR_A}\!\!\!  d\rho (r) \, S_b (\sigma \cap r)+ \lambda \left(1- \int_{\sSA} d\kappa(\sigma) \right)\\
+ \int_{M} \!\! \sqrt{g}\, U \cdot \left(\int_{\sSA} d\kappa(\sigma)\,  d\chi_{J^+(\sigma)} - d\phi \right). \end{split}  \ene
The maximisation of this Lagrangian over $\phi$ gives the U-flow norm bound:
\begin{gather} 
\sup_{\phi} \int_{M} \!\! \sqrt{g} \left (\frac{1}{4G_N}|d\phi \wedgedot d\psi_{\rho}| -
U \cdot d\phi \right ) < \infty  \\[-0.5em] 
\Updownarrow \nonumber \\
U \pm \frac{d\psi_{\rho}}{4G_N} \in j^- \, . \label{eq:maxphi} \end{gather}
To satisfy this constraint, it is necessary for the 1-form $U$ to be past-directed causal.%
\footnote{Note that this is the opposite orientation convention from~\cite{Headrick:2022nbe}. With our convention, the $U$-vector field and its flow lines are future-directed.}

The maximisation over the measure $\kappa$ gives a constraint on the flux of $U$ through all slices:
\begin{gather} 
\sup_{\kappa} \left ( \int_{\sSA} d\kappa \left(-\lambda - \int_\sigma * U + \int_{\sR_A}\!\!\!  d\rho(r) S_{b} (\sigma \cap r)\right)\right) < \infty  \\ 
\Updownarrow \nonumber \\
\forall \sigma\in \sSA:  \lambda \geq  -\int_\sigma * U + \int_{\sR_A}\!\!\!  d\rho (r) S_b (\sigma \cap r).  
\label{eq:Ulaco}
\end{gather}
We have implicitly taken the normal covector on $\sigma$ to be past-directed.

The resulting dual problem is
\bne \begin{split}
&\inf_U \inf_{\lambda} \,(\lambda)
\end{split} \ene 
subject to~\eqref{eq:maxphi} and~\eqref{eq:Ulaco}.

We then eliminate $\lambda$ using the inequality constraint~\eqref{eq:Ulaco} to get the equivalent problem
\bne \label{eq:dua2} \adjustlimits \inf_U \sup_{\sigma\in \sSA} \left ( -\int_\sigma *U + \int_{\sR_A}\!\!\!  d\rho(r) \, S_b (\sigma \cap r)\right) , \quad \text{subject to }U \pm \frac{d\psi_{\rho}}{4G_N} \in j^-\,. \ene

We have that $*U^*|_{D(A) \cup D(A^c)} = 0$ for any optimal $U^*$, because of the norm bound on $U$~\cite{Headrick:2022nbe},
and so we can and will impose $*U|_{D(A) \cup D(A^c)} = 0$ as an additional constraint in the U-flow prescription without affecting the optimum. 

Using Stokes' theorem, we rewrite the flux term $\int_{\sigma} *U$ in~\eqref{eq:dua2} as the divergence $d*U$ integrated over the past $J^-$ of the slice $\sigma$, plus the flux through the past boundary. When $*U|_{D(A) \cup D(A^c)} = 0$, Stokes' theorem implies that
\bne \int_\sigma * U + \int_{J^- (\sigma)}\!\!\! d*U + \int_{\mathcal{I}^-}*U = 0 \,.
\label{eq:Ustokes} \ene
This gives us an integrated divergence term in the dual problem. Next, similarly to the $V$-flow derivation, we will argue that there is guaranteed to exist an optimal U-flow which satisfies $\tilde{\cF}(U^*) = 0$, where
\bne \label{eq:mco3}\tilde{\cF}(U):= \sup_{\sigma\in \sSA} \left (\int_{J^- (\sigma)} \!\!\! d * U + \int_{\sR_A}\!\!\!  d\rho(r) S_b (\sigma \cap r)\right ).
\ene 
The essence of the argument is that, from an optimal flow $U^*$ for the problem~\eqref{eq:dua2} that does not satisfy $\tilde{\cF}(U^*) = 0$, we can construct a new optimal flow $U_{\textrm{new}}^*$ which does.
\begin{enumerate}
\item  First, suppose that we have an optimal U-flow $U^*$ for which $\tilde{\cF}(U^*) >0$. 
Then we need to add more sinks to or remove sources from $J^- (\sigma^*)$.
We attach field lines to past infinity until $\tilde{\cF}(U^*_{\textrm{new}}) = 0$, with the field lines ending at sinks infinitesimally to the future of $\cI^-$. We do not risk violating the norm bound, because if $U^*$ obeys the norm bound then so does $U^* + \delta U$ if $\delta U \in j^-$.

\item Next, suppose that, on the contrary, we have a $U^*$ for which $\tilde{\cF}(U^*) <0$.
Then there must be more sinks than sources for $U^*$ in $J^- (\sigma^*)$, because bulk entropy is non-negative. So, to construct a $U^*_{\rm new}$ satisfying $\tilde{\cF}(U^*_{\rm new})=0$, we need to remove sinks or add sources to $J^- (\sigma^*)$. 
We cannot add sources while staying within the set of optimal flows, because additional sources add to the flux through $\sigma^*$, since the flow is future-oriented. 
Instead, we remove sinks and their attached field lines until $\tilde{\cF}(U_{\rm new}^*) = 0$. 
The new flow, if allowed by the constraints, is still optimal, because we do not change the flux through $\sigma^*$.
Removing sinks risks violating the norm bound, but, since the QES is the only place where the norm bound is saturated for most optimal flows, it is reasonable to assume that there exists an optimal $U^*$ which has a sufficient number of its sinks in $J^- (\sigma^*)$.

\end{enumerate}

We have argued that there is guaranteed to exist an optimal $U^*$ for which $\tilde{\cF}(U^*) = 0$. Then, since we are taking the supremum over slices $\sigma$ in the definition of $\tilde{\cF}$, and because a property obeyed by at least one optimal flow can be imposed as a constraint without affecting the optimum, we can add the following constraint to our dual problem:
\bne \forall \sigma \in \sSA: \left ( \int_{J^- (\sigma)} \!\!\!\!d * U + \int_{\sR_A}\!\!\!  d\rho(r) S_b (\sigma \cap r)\right ) \leq 0 \,.
\ene
To the resulting problem we reintroduce $\rho$-minimisation to finally arrive at our quantum U-flow prescription:
\bne 
\begin{aligned} 
	&\inf_{\rho} \inf_U \left(\int_{\mathcal{I}^-} \!\!\!*U \right) \\
	 \text{subject to:}\quad   &*U|_{D(A) \cup D(A^c)} = 0 \\
	\quad &U \pm \frac{d\psi_{\rho}}{4G_N} \in j^-\\ 
	\quad &\forall \sigma\in \sSA: -\int_{J^-(\sigma)} \!\!\!\!d *U \geq  \int_{\sR_A}\!\!\!  d\rho (r) S_b (\sigma \cap r)\,,
\end{aligned}
\label{eq:qUflow2}
\ene
where $\psi_\rho$ is as defined in~\eqref{eq:phipsidefs}. The measure $\rho$ only appears in the constraints, not in the dual function, so it just has to be feasible. 
This completes our derivation of the quantum U-flow prescription.

Similarly to the V-flow prescription, we can replace the measure $\rho$ and $\psi_{\rho}$ in~\eqref{eq:qUflow2} with a scalar field $\psi$ obeying the ``classical" boundary conditions in~\eqref{Uflow1}, because there exists an optimal $\rho$ with support on a single homology region, which, trivially, is a measure on a nested set of homology regions, so we can require $\rho$ to be a measure on a nested set of homology regions. 

We can also get a prescription that treats the past and future more symmetrically by starting from~\eqref{eq:dua2} and rewriting the flux through $\sigma$ as
\bne 2 \int_\sigma *U = \int_{J^+ (\sigma)}\!\!\!\!\!\! d*U + \int_{\mathcal{I}^+}\!\!\!\! *U - \int_{J^- (\sigma)}\!\!\!\! d*U  - \int_{\mathcal{I}^-}\!\!\!\! *U \,, \ene
and this leads to the dual problem
\bne \begin{split} 
&\frac{1}{2} \inf_U \left( \int_{\mathcal{I}^-} \!\!\!\!*U - \int_{\cI^+}\!\!\! * U \right) \\
\text{subject to } &*U|_{D(A) \cup D(A^c)} = 0\\ 
 &U \pm \frac{d\psi_{\rho}}{4G_N} \in j^-\\ 
\forall \sigma\in \sSA: &\int_{J^+(\sigma)}\!\!\!\! d *U - \int_{J^-(\sigma)}\!\!\!\! d *U \,\geq \,  2\int_{\sR_A}\!\!\!  d\rho (r) S_b (\sigma \cap r) \,. \end{split} \ene

\acknowledgments

AR would like to thank Juan Pedraza and Andy Svesko for initial collaboration. The work of MH was supported by the U.S.\ Department of Energy through award DE-SC0009986, and by a Simons Sabbatical Visiting Chair at the Institut des Hautes \'Etudes Scientifiques. The work of SRK is supported in part by ISF grant no. 2159/22, by Simons Foundation grant 994296 (Simons Collaboration on Confinement and QCD Strings), by the Minerva foundation with funding from the Federal
German Ministry for Education and Research, and by the German Research Foundation
through a German-Israeli Project Cooperation (DIP) grant “Holography and the Swampland”. The work of AR is supported by FWO-Vlaanderen through a Senior Postdoctoral Fellowship 1223125N.

\appendix

\section{Failure of level-set-relaxed quantum maximin to be concave}
\label{sec:noc}

We will show that the bulk entropy component of the objective of the level-set-relaxed quantum maximin prescription, Eq.~\eqref{eq:ppqmm}, is not concave in $\phi$. This component is the averaged bulk entropy
\bne \overline{\Sb}(\phi,\psi):= \int_0^1 dt \int_0^1 ds S_{\textrm{b}} (\sigma_t \cap r_s) ,\ene
Thus, the level set/scalar function approach to relaxing quantum maximin is not a concave optimisation problem in $\phi$, which blocks us from dualising to an equivalent U-flow prescription, because the optimal duality gap is not guaranteed to vanish.

In contrast, there is no obstacle to dualising on $\psi$, because the level-set-relaxed quantum maximin prescription \textit{is} convex in $\psi$. The prescription is convex in $\psi$ because both of its components are; the averaged entropy $\overline{\Sb}$ is, as can be shown using strong subadditivity of the bulk entropies~\cite{Headrick:2025awv}, and so too the area term $|d\phi \wedgedot d\psi|$~\cite{Headrick:2022nbe}.

\begin{figure}
    \centering
    \begin{overpic}[width=0.95\linewidth]{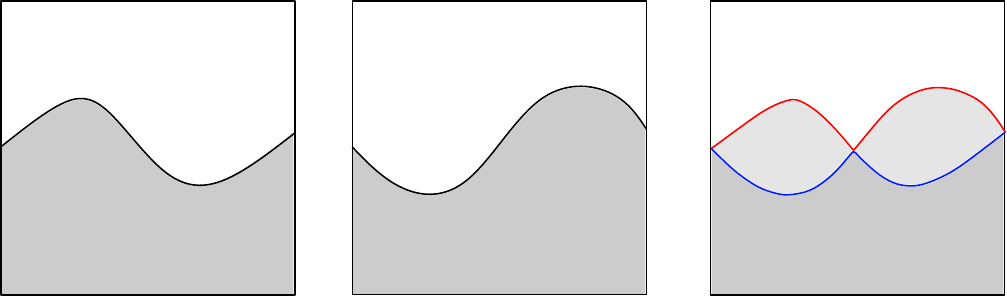}
    \put(10,25){$\phi_1 =1$}
    \put(10,3){$\phi_1 =0$}
    \put(18,12){$\sigma_1$}
    \put(45,25){$\phi_2 =1$}
    \put(45,3){$\phi_2 =0$}
    \put(41,11){$\sigma_2$}
    \put(41,11){$\sigma_2$}
    \put(71.5,18.5){\textcolor{red}{$\sigma_+$}}
    \put(72,10){\textcolor{blue}{$\sigma_-$}}
    \put(82,3){$\phi_3 = 0$}
    \put(82,25){$\phi_3 = 1$}
    \put(77,14){$\frac{1}{2}$}
    \put(91,15){$\frac{1}{2}$}
    \end{overpic}
    \caption{The cross-section (on a timesheet) of two slices $\sigma_1$ and $\sigma_2$, and two slices $\sigma_\pm$ constructed from them. The shaded regions represent the corresponding level set functions $\phi_i$ and their values.}
    \label{fig:cc2}
\end{figure}

For $\overline{\Sb}$ to be concave in $\phi$ would require, for all pairs of $\phi_1$ and $\phi_2$ (suppressing $\psi$-dependence), 
\bne \overline{\Sb}(\alpha \phi_1 + (1-\alpha)\phi_2) \geq \alpha \overline{\Sb}(\phi_1) + (1-\alpha)\overline{\Sb} (\phi_2) \label{eq:concav_rel} \ene
for $\alpha \in [0,1]$. We will construct a counterexample to this.

Consider a pair of slices $\sigma_1$ and $\sigma_2$ which are not causally ordered ($J^\pm (\sigma_1) \nsubseteq J^\pm (\sigma_2)$). 
Let us define a pair of scalar functions
\bne \phi_i(x) = \begin{cases} &0, \quad x \in J^- (\sigma_i) \\
&1, \quad \text{otherwise.} \end{cases} \label{eq:phiidef}\ene
The derivatives $d\phi_i$ are delta-function distributions localised on $\sigma_i$ and antiparallel to the slice's normal covector.
From these $\phi_i$, we define $\phi_3 := \frac{1}{2}(\phi_1 + \phi_2)$, which takes three different values, $\phi_3 \in \{ 0,1/2,1\}$, and the boundary of these level sets defines two new slices $\sigma_\pm$. 
The slices are causally ordered: $J^+ (\sigma_+) \subseteq J^+ (\sigma_-)$. 
In Fig.~\ref{fig:cc2}, we illustrate the intersection of these four slices with an arbitrary fixed timesheet $\tau(r)$.

Taking these $\phi_i$, and $\alpha = 1/2$, the concavity relation~\eqref{eq:concav_rel} becomes
\bne \label{eq:CCW2}  S_{\textrm{b}}(\sigma_+ \cap r) + S_{\textrm{b}}(\sigma_- \cap r) \geq S_{\textrm{b}}(\sigma_1 \cap r) + S_{\textrm{b}}(\sigma_2 \cap r). \ene
The following bulk state is a counterexample to this inequality: a Bell pair of particles that pass through the timesheet, with one partner passing through the left-hand $\phi_3 = \frac{1}{2}$ region in Fig.~\ref{fig:cc2}, and the other through the right-hand $\phi_3 = \frac{1}{2}$ region. It is a counterexample because the Bell pair contributes positively to the entropies $\Sb ( \sigma_1 \cap r)$ and $\Sb (\sigma_2 \cap r)$, and nothing to $\Sb (\sigma_+ \cap r)$ and $\Sb (\sigma_- \cap r)$. Note that Eq.~\eqref{eq:CCW2} would still not be true if we were to replace $S_{\textrm{b}}$ with $S_{\textrm{gen}}$, because the resulting area terms on the two sides of the inequality cancel.

\bibliographystyle{JHEP}
\bibliography{biblio.bib}

@article{Harper:2021uuq,
    author = "Harper, Jonathan",
    title = "{Hyperthreads in holographic spacetimes}",
    eprint = "2107.10276",
    archivePrefix = "arXiv",
    primaryClass = "hep-th",
    reportNumber = "BRX-TH-6688",
    doi = "10.1007/JHEP09(2021)118",
    journal = "JHEP",
    volume = "09",
    pages = "118",
    year = "2021"
}

@article{Harper:2019lff,
    author = "Harper, Jonathan and Headrick, Matthew",
    title = "{Bit threads and holographic entanglement of purification}",
    eprint = "1906.05970",
    archivePrefix = "arXiv",
    primaryClass = "hep-th",
    doi = "10.1007/JHEP08(2019)101",
    journal = "JHEP",
    volume = "08",
    pages = "101",
    year = "2019"
}

@article{Du:2024xoz,
    author = "Du, Dong-Hui and Sun, Jia-Rui",
    title = "{Towards bit threads in general gravitational spacetimes}",
    eprint = "2406.04092",
    archivePrefix = "arXiv",
    primaryClass = "hep-th",
    doi = "10.1007/JHEP08(2024)016",
    journal = "JHEP",
    volume = "08",
    pages = "016",
    year = "2024"
}

@article{Headrick:2020gyq,
    author = "Headrick, Matthew and Held, Jesse and Herman, Joel",
    title = "{Crossing Versus Locking: Bit Threads and Continuum Multiflows}",
    eprint = "2008.03197",
    archivePrefix = "arXiv",
    primaryClass = "hep-th",
    reportNumber = "BRX-TH-6667",
    doi = "10.1007/s00220-022-04476-w",
    journal = "Commun. Math. Phys.",
    volume = "396",
    number = "1",
    pages = "265--313",
    year = "2022"
}

@article{lieb1973proof,
  title={Proof of the strong subadditivity of quantum-mechanical entropy},
  author={Lieb, Elliott H and Ruskai, Mary Beth},
  journal={Les rencontres physiciens-math{\'e}maticiens de Strasbourg-RCP25},
  volume={19},
  pages={36--55},
  year={1973}
}

@article{Harper:2022sky,
    author = "Harper, Jonathan",
    title = "{Perfect tensor hyperthreads}",
    eprint = "2205.01140",
    archivePrefix = "arXiv",
    primaryClass = "hep-th",
    reportNumber = "BRX-TH-6705",
    doi = "10.1007/JHEP09(2022)239",
    journal = "JHEP",
    volume = "09",
    pages = "239",
    year = "2022"
}

@article{Gupta:2025jlq,
    author = "Gupta, Divij and Headrick, Matthew and Sasieta, Martin",
    title = "{Entangled universes}",
    eprint = "2505.08945",
    archivePrefix = "arXiv",
    primaryClass = "hep-th",
    doi = "10.21468/SciPostPhys.19.6.143",
    journal = "SciPost Phys.",
    volume = "19",
    number = "6",
    pages = "143",
    year = "2025"
}

@article{Ryu:2006bv,
    author = "Ryu, Shinsei and Takayanagi, Tadashi",
    title = "{Holographic derivation of entanglement entropy from AdS/CFT}",
    eprint = "hep-th/0603001",
    archivePrefix = "arXiv",
    reportNumber = "NSF-KITP-06-11",
    doi = "10.1103/PhysRevLett.96.181602",
    journal = "Phys. Rev. Lett.",
    volume = "96",
    pages = "181602",
    year = "2006"
}

@article{Hubeny:2007xt,
    author = "Hubeny, Veronika E. and Rangamani, Mukund and Takayanagi, Tadashi",
    title = "{A Covariant holographic entanglement entropy proposal}",
    eprint = "0705.0016",
    archivePrefix = "arXiv",
    primaryClass = "hep-th",
    reportNumber = "DCPT-07-13, KUNS-2069",
    doi = "10.1088/1126-6708/2007/07/062",
    journal = "JHEP",
    volume = "07",
    pages = "062",
    year = "2007"
}

@article{Freedman:2016zud,
    author = "Freedman, Michael and Headrick, Matthew",
    title = "{Bit threads and holographic entanglement}",
    eprint = "1604.00354",
    archivePrefix = "arXiv",
    primaryClass = "hep-th",
    reportNumber = "BRX-TH-6302, NSF-KITP-16-051",
    doi = "10.1007/s00220-016-2796-3",
    journal = "Commun. Math. Phys.",
    volume = "352",
    number = "1",
    pages = "407--438",
    year = "2017"
}

@article{Rolph:2021hgz,
    author = "Rolph, Andrew",
    title = "{Quantum bit threads}",
    eprint = "2105.08072",
    archivePrefix = "arXiv",
    primaryClass = "hep-th",
    doi = "10.21468/SciPostPhys.14.5.097",
    journal = "SciPost Phys.",
    volume = "14",
    number = "5",
    pages = "097",
    year = "2023"
}

@article{Headrick:2022nbe,
    author = "Headrick, Matthew and Hubeny, Veronika E.",
    title = "{Covariant bit threads}",
    eprint = "2208.10507",
    archivePrefix = "arXiv",
    primaryClass = "hep-th",
    doi = "10.1007/JHEP07(2023)180",
    journal = "JHEP",
    volume = "07",
    pages = "180",
    year = "2023"
}

@article{Engelhardt:2014gca,
    author = "Engelhardt, Netta and Wall, Aron C.",
    title = "{Quantum Extremal Surfaces: Holographic Entanglement Entropy beyond the Classical Regime}",
    eprint = "1408.3203",
    archivePrefix = "arXiv",
    primaryClass = "hep-th",
    doi = "10.1007/JHEP01(2015)073",
    journal = "JHEP",
    volume = "01",
    pages = "073",
    year = "2015"
}

@article{Akers:2019lzs,
    author = "Akers, Chris and Engelhardt, Netta and Penington, Geoff and Usatyuk, Mykhaylo",
    title = "{Quantum Maximin Surfaces}",
    eprint = "1912.02799",
    archivePrefix = "arXiv",
    primaryClass = "hep-th",
    doi = "10.1007/JHEP08(2020)140",
    journal = "JHEP",
    volume = "08",
    pages = "140",
    year = "2020"
}

@article{Almheiri:2019psf,
    author = "Almheiri, Ahmed and Engelhardt, Netta and Marolf, Donald and Maxfield, Henry",
    title = "{The entropy of bulk quantum fields and the entanglement wedge of an evaporating black hole}",
    eprint = "1905.08762",
    archivePrefix = "arXiv",
    primaryClass = "hep-th",
    doi = "10.1007/JHEP12(2019)063",
    journal = "JHEP",
    volume = "12",
    pages = "063",
    year = "2019"
}

@article{Penington:2019npb,
    author = "Penington, Geoffrey",
    title = "{Entanglement Wedge Reconstruction and the Information Paradox}",
    eprint = "1905.08255",
    archivePrefix = "arXiv",
    primaryClass = "hep-th",
    doi = "10.1007/JHEP09(2020)002",
    journal = "JHEP",
    volume = "09",
    pages = "002",
    year = "2020"
}

@article{sion1958general,
  title={On general minimax theorems.},
  author={Sion, Maurice},
  year={1958},
  journal= "Pacific Journal of Mathematics",
  volume="8(1)",
  pages="171-176"
}

@article{Almheiri:2019qdq,
    author = "Almheiri, Ahmed and Hartman, Thomas and Maldacena, Juan and Shaghoulian, Edgar and Tajdini, Amirhossein",
    title = "{Replica Wormholes and the Entropy of Hawking Radiation}",
    eprint = "1911.12333",
    archivePrefix = "arXiv",
    primaryClass = "hep-th",
    doi = "10.1007/JHEP05(2020)013",
    journal = "JHEP",
    volume = "05",
    pages = "013",
    year = "2020"
}

@article{Agon:2021tia,
    author = "Ag{\'o}n, Cesar A. and Pedraza, Juan F.",
    title = "{Quantum bit threads and holographic entanglement}",
    eprint = "2105.08063",
    archivePrefix = "arXiv",
    primaryClass = "hep-th",
    reportNumber = "BRX-TH-6679",
    doi = "10.1007/JHEP02(2022)180",
    journal = "JHEP",
    volume = "02",
    pages = "180",
    year = "2022"
}

@article{Headrick:2025awv,
    author = "Headrick, Matthew and Kasireddy, Sreeman Reddy and Rolph, Andrew",
    title = "{Quantum bit threads and the entropohedron}",
    eprint = "2510.22601",
    archivePrefix = "arXiv",
    primaryClass = "hep-th",
    doi = "10.1007/JHEP04(2026)196",
    journal = "JHEP",
    volume = "04",
    pages = "196",
    year = "2026"
}

@article{Harper:2018sdd,
    author = "Harper, Jonathan and Headrick, Matthew and Rolph, Andrew",
    title = "{Bit Threads in Higher Curvature Gravity}",
    eprint = "1807.04294",
    archivePrefix = "arXiv",
    primaryClass = "hep-th",
    reportNumber = "BRX-TH-6327, MIT-CTP/500, MIT-CTP-500",
    doi = "10.1007/JHEP11(2018)168",
    journal = "JHEP",
    volume = "11",
    pages = "168",
    year = "2018"
}

@article{Camps:2013zua,
    author = "Camps, Joan",
    title = "{Generalized entropy and higher derivative Gravity}",
    eprint = "1310.6659",
    archivePrefix = "arXiv",
    primaryClass = "hep-th",
    doi = "10.1007/JHEP03(2014)070",
    journal = "JHEP",
    volume = "03",
    pages = "070",
    year = "2014"
}

@article{Dong:2013qoa,
    author = "Dong, Xi",
    title = "{Holographic Entanglement Entropy for General Higher Derivative Gravity}",
    eprint = "1310.5713",
    archivePrefix = "arXiv",
    primaryClass = "hep-th",
    reportNumber = "SU-ITP-13-21",
    doi = "10.1007/JHEP01(2014)044",
    journal = "JHEP",
    volume = "01",
    pages = "044",
    year = "2014"
}

@book{Boyd:2004fnq,
    author = "Boyd, Stephen and Vandenberghe, Lieven",
    title = "{Convex Optimization}",
    doi = "10.1017/cbo9780511804441",
    isbn = "978-0-521-83378-3, 978-0-511-80444-1",
    publisher = "Cambridge University Press",
    month = "3",
    year = "2004"
}

\end{document}